\documentclass[aps,prc,preprint,superscriptaddress]{revtex4-1}

\usepackage[english]{babel}
\usepackage[utf8x]{inputenc}
\usepackage[T1]{fontenc}

\usepackage{color}
\usepackage{xspace}
\usepackage{adjustbox}
\usepackage{relsize}
\usepackage{booktabs}
\usepackage{enumitem}

\usepackage[a4paper,top=3cm,bottom=2cm,left=3cm,right=3cm,marginparwidth=1.75cm]{geometry}

\usepackage{amsmath}
\usepackage{graphicx}
\usepackage[colorlinks=true, allcolors=blue]{hyperref}

\def\pp{pp\xspace}
\def\pA{pA\xspace}
\def\AA{AA\xspace}
\def\pPb{p--Pb\xspace}
\def\PbPb{Pb--Pb\xspace}
\def\AuAu{Au--Au\xspace}
\let\mrm\mathrm

\def\GeV{{\ensuremath\mathrm{Ge\kern-.07emV}}\xspace}
\def\TeV{{\ensuremath\mathrm{T\kern-.05em e\kern-.07emV}}\xspace}
\def\GeVc{{\ensuremath\GeV\kern-.2em/\kern-.05em\mathrm{c}}\xspace}
\def\pT{{\ensuremath p_{\mathrm{T}}}\xspace}
\def\ET{{\ensuremath E_{\mathrm{T}}}\xspace}
\def\nsNN{{\ensuremath\sqrt{s_{\scriptscriptstyle\mathrm{NN}}}}\xspace}
\def\sNN#1#2{{\ensuremath\nsNN=#1\,\mathrm{#2}}\xspace}
\def\RAA{{\ensuremath R_{AA}}\xspace}
\def\rivet{\textsc{Rivet}\xspace}
\def\hepmc{\textsc{HepMC}\xspace}
\def\pythia{\textsc{Pythia}~8.2\xspace}
\def\angantyr{\pythia/\textsc{Angantyr}\xspace}
\def\yoda{\textsc{Yoda}\xspace}
\def\alice{\textsc{Alice}\xspace}
\def\atlas{\textsc{Atlas}\xspace}
\def\cms{\textsc{Cms}\xspace}
\def\brahms{\textsc{Brahms}\xspace}
\def\star{\textsc{Star}\xspace}
\def\jewel{\textsc{Jewel}\xspace}

\def\eposlhc{\textsc{Epos-lhc}\xspace}
\def\lhc{\textsc{Lhc}\xspace}
\def\mcplots{\textsc{MCplots}\xspace}
\def\herwigpista{\textsc{Herwig}+\textsc{Pista}\xspace}
\def\jetscape{\textsc{Jetscape}\xspace}
\def\mc#1{Monte--Carlo\xspace #1}
\newcommand{\Cxx}{\hbox{C\raisebox{0.2ex}{\smaller \kern-.1em+\kern-.2em+}}}
\def\twofigw{.49\linewidth}
\def\threefigw{.33\linewidth}

\begin{document}

\title{Confronting Experimental Data with Heavy-Ion Models \\[1ex]
  \large \rivet~for Heavy Ions}

\author{Christian~Bierlich}
\affiliation{Niels Bohr Institute, Blegdamsvej 17, 2100 Copenhagen, Denmark}
\affiliation{Dept. of Astronomy and Theoretical Physics, Lund University, S{\"o}lvegatan
  14A, S-223 62 Lund, Sweden}
\author{Andy~Buckley}
\affiliation{School of Physics and Astronomy, University of Glasgow, G12
  8QQ, Glasgow, UK}
\author{Christian Holm Christensen}
\affiliation{Niels Bohr Institute, Blegdamsvej 17, 2100 Copenhagen, Denmark}
\author{Peter Harald Lindenov Christiansen}
\affiliation{Dept. of Physics, Lund University, Professorsgatan 1, S-223 62 Lund, Sweden}
\author{Cody~B.~Duncan}
\affiliation{School of Physics and Astronomy, Monash University, Clayton,
  VIC 3800, Australia}
\affiliation{Institute for Theoretical Physics, Karlsruhe Institut f\"ur
  Technologie, Wolfgang-Gaede-Stra\ss e 1, 76131 Karlsruhe Germany}
\author{Jan~Fiete~Grosse-Oetringhaus}
\affiliation{CERN, Esplanade de Particules 1, Geneva, Switzerland}
\author{Przemyslaw~Karczmarczyk}
\affiliation{Faculty of Physics, Warsaw University of Technology,
  Koszykowa 75, 00-662 Warszawa, Poland}
\affiliation{CERN, Esplanade de Particules 1, Geneva, Switzerland}
\author{Patrick~Kirchgae\ss er}
\affiliation{Institute for Theoretical Physics, Karlsruhe Institut f\"ur
  Technologie, Wolfgang-Gaede-Stra\ss e 1, 76131 Karlsruhe Germany}
\author{Jochen~Klein}
\affiliation{Istituto Nazionale di Fisica Nucleare, Sezione di Torino,
  Italy}
\affiliation{CERN, Esplanade de Particules 1, Geneva, Switzerland}
\author{Leif~L\"onnblad}
\affiliation{Dept. of Astronomy and Theoretical Physics, Lund University, S{\"o}lvegatan
  14A, S-223 62 Lund, Sweden}
\author{Roberto~Preghenella}
\affiliation{Istituto Nazionale di Fisica Nucleare, Sezione di Bologna,
  Italy}
\author{Christine~O.~Rasmussen}
\affiliation{Dept. of Astronomy and Theoretical Physics, Lund University, S{\"o}lvegatan
  14A, S-223 62 Lund, Sweden}
\author{Maria~Stefaniak}
\affiliation{Faculty of Physics, Warsaw University of Technology,
  Koszykowa 75, 00-662 Warszawa, Poland}
\affiliation{Subatech -- IMT Atlantique, 4 rue Alfred Kastler, 44307
  Nantes, France}
\author{Vytautas~Vislavicus}
\affiliation{Niels Bohr Institute, Blegdamsvej 17, 2100 Copenhagen, Denmark}

%\author[1,2]{Christian~Bierlich}
%\author[3]{Andy~Buckley}
%\author[2]{Christian Holm Christensen}
%\author[1]{Peter Harald Lindenov Christiansen}
%\author[4,5]{Cody~B.~Duncan}
%\author[10]{Jan~Fiete~Grosse-Oetringhaus}
%\author[8,10]{Przemyslaw~Karczmarczyk}
%\author[5]{Patrick~Kirchgae\ss er}
%\author[9,10]{Jochen~Klein}
%\author[1]{Leif~L\"onnblad}
%\author[6]{Roberto~Preghenella}
%\author[1]{Christine~O.~Rasmussen}
%\author[7,8]{Maria~Stefaniak}
%\author[2]{Vytautas~Vislavicus}

%\affil[1]{Dept. of Astronomy and Theoretical Physics, S{\" o}lvegatan
%  14A, S-223 62 Lund, Sweden}
%\affil[2]{Niels Bohr Institute, Blegdamsvej 17, 2100 Copenhagen, Denmark}
%\affil[3]{School of Physics and Astronomy, University of Glasgow, G12
%  8QQ, Glasgow, UK}
%\affil[4]{School of Physics and Astronomy, Monash University, Clayton,
%  VIC 3800, Australia}
%\affil[5]{Institute for Theoretical Physics, Karlsruhe Institut f\"ur
%  Technologie, Wolfgang-Gaede-Stra\ss e 1, 76131 Karlsruhe Germany}
%\affil[6]{Istituto Nazionale di Fisica Nucleare, Sezione di Bologna,
%  Italy}
%\affil[7]{Subatech -- IMT Atlantique, 4 rue Alfred Kastler, 44307
%  Nantes, France}
%\affil[8]{Faculty of Physics, Warsaw University of Technology,
%  Koszykowa 75, 00-662 Warszawa, Poland}
%\affil[9]{Istituto Nazionale di Fisica Nucleare, Sezione di Torino,
%  Italy}
%\affil[10]{CERN, Geneva, Switzerland}

\date{\today}
\preprint{LU TP 20-04}
\preprint{MCNET-20-04}
\begin{abstract}
  The \rivet~library is an important toolkit in particle physics, and
  serves as a repository for analysis data and code. It allows for
  comparisons between data and theoretical calculations of the final
  state of collision events. This paper outlines several recent
  additions and improvements to the framework to include support for
  analysis of heavy ion collision simulated data.  The paper also
  presents examples of these recent developments and their
  applicability in implementing concrete physics analyses.
\end{abstract}

\maketitle

\section{Introduction}

High energy collisions of hadrons have, with access to data from
colliders at the energy frontier over the past 20 years, been the main
avenue for precision studies of the phenomenology of the strong
nuclear force (QCD). Since such collisions involve a vast amount of
different phenomena --- some calculable from first principles, other
relying on models --- development has tended towards large calculation
packages, which attempt to include as many effects as possible, with
the aim of simulating full ``events'' resembling the collisions
observed by experiments.  While such event generators are very useful,
their individual model components are difficult to validate in a
systematic way.  This is because adding new model components may
compromise existing agreement with some results while improving
agreement with other results.  To overcome this challenge, mass
comparison with current and past data is needed. Such comparisons are,
for the most widespread proton--proton (\pp) event generators,
facilitated by the software package \rivet~\cite{Buckley:2010ar}.

Heavy-ion physics has entered the precision era and quantitative and
comprehensive model comparisons have become
crucial~\cite{Schukraft:2017nbn}.  Technical difficulties --- similar
to those faced earlier for collisions of protons --- now also surface
in this area of research. Inclusion of several new model components on
the theoretical side, for example in the \jetscape generator
package~\cite{Putschke:2019yrg}, introduces similar difficulties of
validation, well known from hadron collisions. Furthermore, the
discovery of QGP--like behaviour in collisions of small systems,
i.e. \pp and proton--ion
(\pA)~\cite{Khachatryan:2010gv,ALICE:2017jyt}, suggests that similar
physics is at play in these collisions. These points strengthen the
need for a generalized and common method of performing systematic
comparisons of theoretical models to data for proton--proton,
nucleon--nucleus, and ion--ion (\AA) collisions.  Such a method will
not only be practically useful but would also enable new avenues for
discovering similarities between two previously quite detached fields
of subatomic physics~\cite{Citron:2018lsq}.

This paper introduces the new features of the \rivet framework to
allow one to obtain a direct comparison between \mc{event generators} and experimental
results from \pA and \AA collisions. In Section~\ref{sec:rivet} a
brief introduction to the framework is given, along with additions
which are motivated by requirements of heavy ion analyses, but are of
more general nature.  In Section~\ref{sec:features} two further
additions, more specialized to specific heavy ion analyses are
introduced: 1) estimation of centrality in
Section~\ref{sec:centrality}) and 2) the Generic Framework for flow
observables in Section \ref{sec:generic-framework}. In
Section~\ref{sec:example}, a simple analysis example is given, with
guidelines for the user to run the analysis and obtain a simple
comparison. For the two latter additions, Section~\ref{sec:biases}
presents use-cases where the need for calculating final-state
observables in the same way as the experiment are obvious.  For
centrality estimation (in Section~\ref{sec:cent-bias}) we show how
different centrality estimators on final state and initial state level
change results on basic quantities like multiplicity at mid-rapidity,
in particular in \pA, but to some degree also in \AA. For flow
observables (in Section~\ref{sec:non-flow}) we show how important
non-flow contributions are captured using the generic framework
implementation, as opposed to a simple calculation based on the
initial state event plane.

\section{The \rivet~framework and new features of general nature}
\label{sec:rivet}

\rivet~is a computational framework\footnote{A brief review of the
  framework, with emphasis on functionality necessary for later
  sections, is given here.  For a full description of the
  \rivet~framework, the reader is referred to the original manual
  \cite{Buckley:2010ar} or the manual of \rivet~version 3
  \cite{Bierlich:2019rhm}.}, written as a \Cxx{} shared library
supplying core functionality, and a large number (949 at the time of
writing) of \Cxx{} plugin libraries implementing collider analysis
routines.

\begin{figure}
  \begin{center}
    \includegraphics[width=0.7\linewidth]{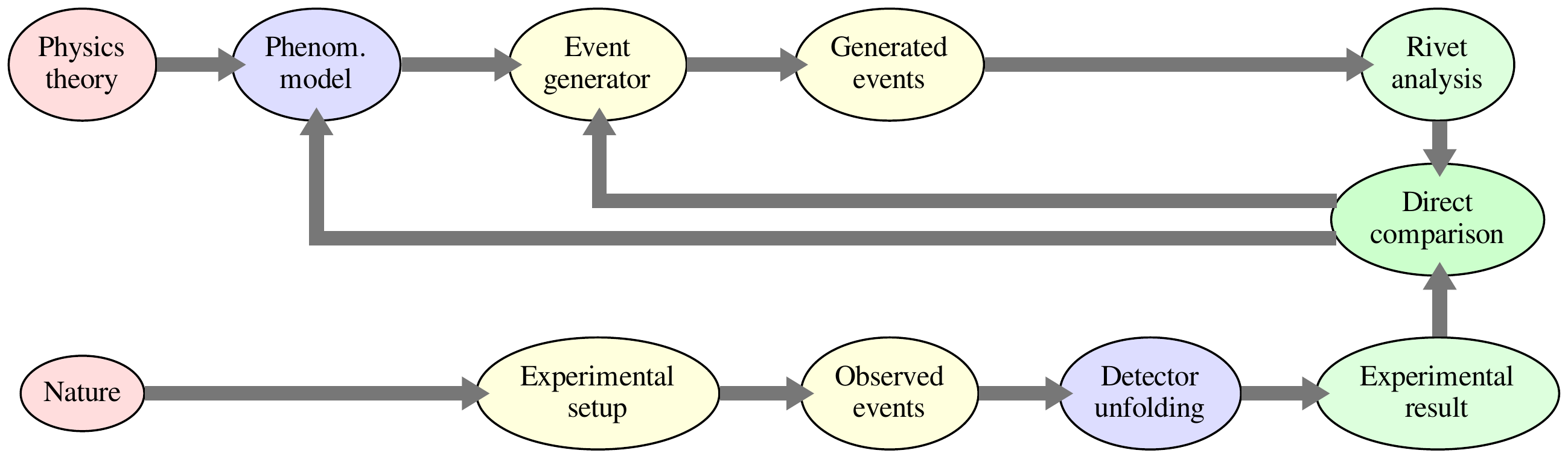}
  \end{center}
  \caption{\label{fig:rivet}Outline of how \rivet~connects
    experimental analyses to theory and validation, with the typical
    workflow of a physics program shown. On the top row, a typical
    theoretical workflow is depicted, where physics theory informs a
    phenomenological model implemented in an event generator.  On the
    bottom row, the experimental workflow is shown in a similar
    fashion.}
\end{figure}

In Figure~\ref{fig:rivet}, the \rivet~frameworks' connection to
experiment and theory is outlined.  The figure illustrates how
comparison of experimental results to results from event generators
provide a feed-back loop to the phenomenological models and the
concrete event generator development. This type of feedback loop is
widely used for development of \mc{event
  generators}~\cite{Sjostrand:2014zea,Bellm:2015jjp,Biro:2019ijx}, as
well as their validation and
tuning~\cite{Skands:2014pea,Khachatryan:2015pea,ATLAS:2012uec}.

\rivet analyses are performed as comparisons to data unfolded to
``particle level'', meaning that the experimental data should be
processed to a level where it is directly comparable to \mc output, as
a list of stable or unstable particles.

After a short technical introduction in Section~\ref{sec:technical},
the following sections present additions to \rivet motivated by heavy
ion development, but of a general nature.  In
Section~\ref{sec:primary-particle} an example of implementing
experimental particle definitions is shown.  In
Section~\ref{sec:reentrant} the ability to generate cross-system
observables by running the same analysis on \mc for both \AA/\pA and
\pp, followed by the ability to run analyses in different ``modes''
specified by analysis options in
Section~\ref{sec:analysis-options}. Finally, the possibility to
perform event mixing is presented in Section~\ref{sec:event-mixing}.

\subsection{Technical introduction}
\label{sec:technical}

The core \rivet library provides tools to process simulated events in
the \hepmc format~\cite{Dobbs:2001ck} with the aim of calculating
observable quantities.  The \texttt{Projection} paradigm is one of the
key \rivet concepts. Each projection calculates a specific observable
from the full, generated event. This can, for example, be all charged
particles in a certain fiducial volume, all possible reconstructed
$Z$-bosons decaying leptonically, event shapes, or, as introduced in
the Section~\ref{sec:features}, event centralities or
$Q$-vectors. There are two key advantages to projections. The first is
computational.  Since a projection can be reused across analyses,
several analyses can be run simultaneously without the need to repeat
a ``particle loop'' for each analysis. Secondly, and most importantly,
projections serve as convenient shorthands when implementing an
experimental analysis. There is no need to repeat the implementation
to select stable particles, or of a certain experimental trigger
definition if it already exists as a pre-defined projection.

\begin{figure}[htbp]
  \centering
  \includegraphics[width=.7\linewidth]{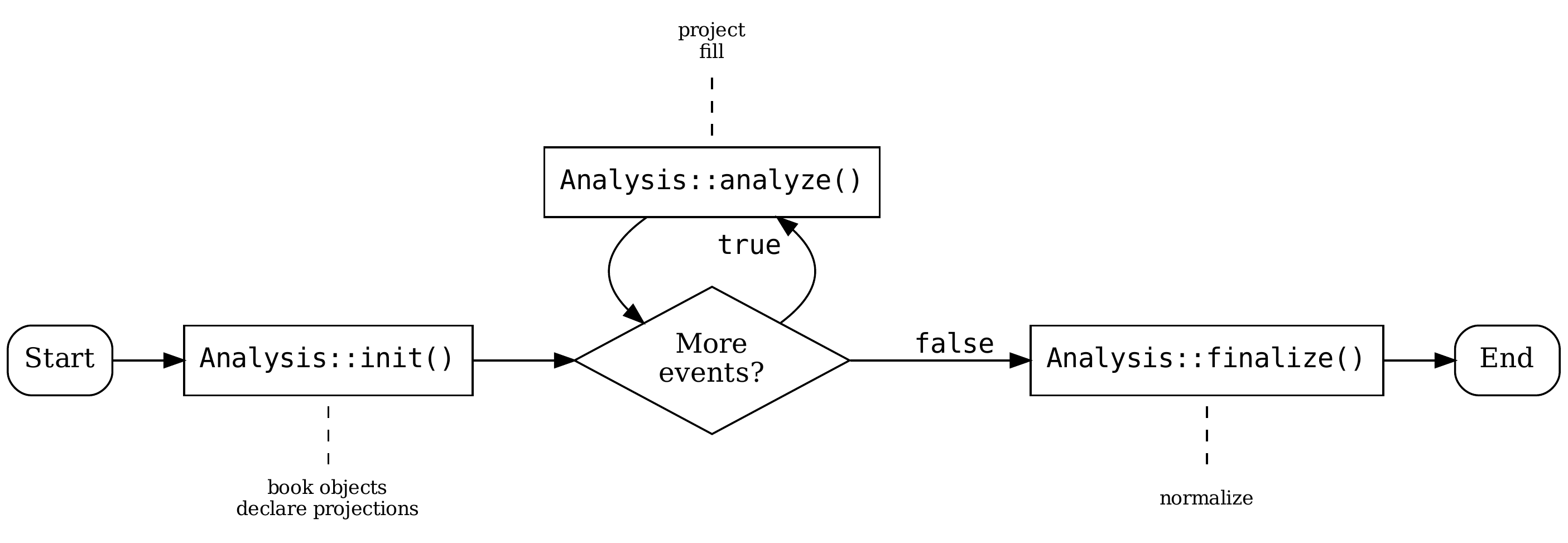}
  \caption{Execution flow of an \texttt{Analysis} class.  }
  \label{fig:anaflow}
\end{figure}
A \rivet analyses is split into three parts, implemented as separate
member functions of an analysis class, inheriting from the
\texttt{Analysis} base class

\begin{description}
\item[Initialization]: The \texttt{init()} method is called once in
  the beginning of the analysis, and is intented for the user to
  initialize (or ``book'') objects such as projections or histograms
  used throughout the analysis.
\item[Analysis]: The \texttt{analyze(const Rivet::Event\&)} method is
  called once per event, and is intended for the user to implement
  calculation of single event quantities, and fill histograms.
\item[Finalization]: The \texttt{finalize()} method is called once at
  the end of the analysis, and is intended for the user to apply the
  correct normalization of histograms, compute all-event averages and
  possibly construct ratios of aggregated quantities. In
  Section~\ref{sec:reentrant} the possibility to run this method for a
  merged sample is outlined.
\end{description}

\subsection{The \hepmc~heavy ion container}

A \rivet analysis has access to heavy-ion specific generator
information, through the heavy-ion container of a \hepmc input event,
provided that it is filled by the event generator. The heavy-ion
container contains several data members, which in general are model
dependent generation information, in most cases linked to the Glauber
model \cite{Glauber:1955qq,Miller:2007ri}, such as number of wounded
nucleons, number of hard sub-collisions and number of spectator
neutrons/protons. The heavy-ion record also holds more general
information, such as the impact parameter, the event plane angle, and
from \hepmc version 2.6.10 onward, it is possible to set a value for
generator-level centrality.  Detailed information about data members
available in \rivet, and for implementing a \hepmc heavy-ion container
in an event generator, is provided in the projects' respective online
documentation.  It is important to note the following for analyses
implemented using information from the \hepmc heavy-ion container:

\begin{itemize}
\item There is no guarantee that an event generator implements the
  heavy-ion container, and one should perform appropriate tests to
  avoid code failure.
\item There is no guarantee that the generator definition of a certain
  model dependent quantity is equivalent to the same as used in the
  experimental analysis.
\end{itemize}

\subsection{Implementing experimental primary-particle definitions}
\label{sec:primary-particle}

% The full phase space output from a Monte Carlo event generator, cannot
% be compared directly to data, due to limited acceptance of physical
% detectors\footnote{In so far as the generator \emph{does} provide
%   this.  Some generators provide a \emph{limited} phase-space event,
%   and the remarks above apply equally in that case.}. The \rivet
% \texttt{Projections} introduced above can be used to reduce the
% particles to the acceptance of the considered detector.

\begin{figure}
  \begin{center}
    \includegraphics[width=0.4\linewidth]{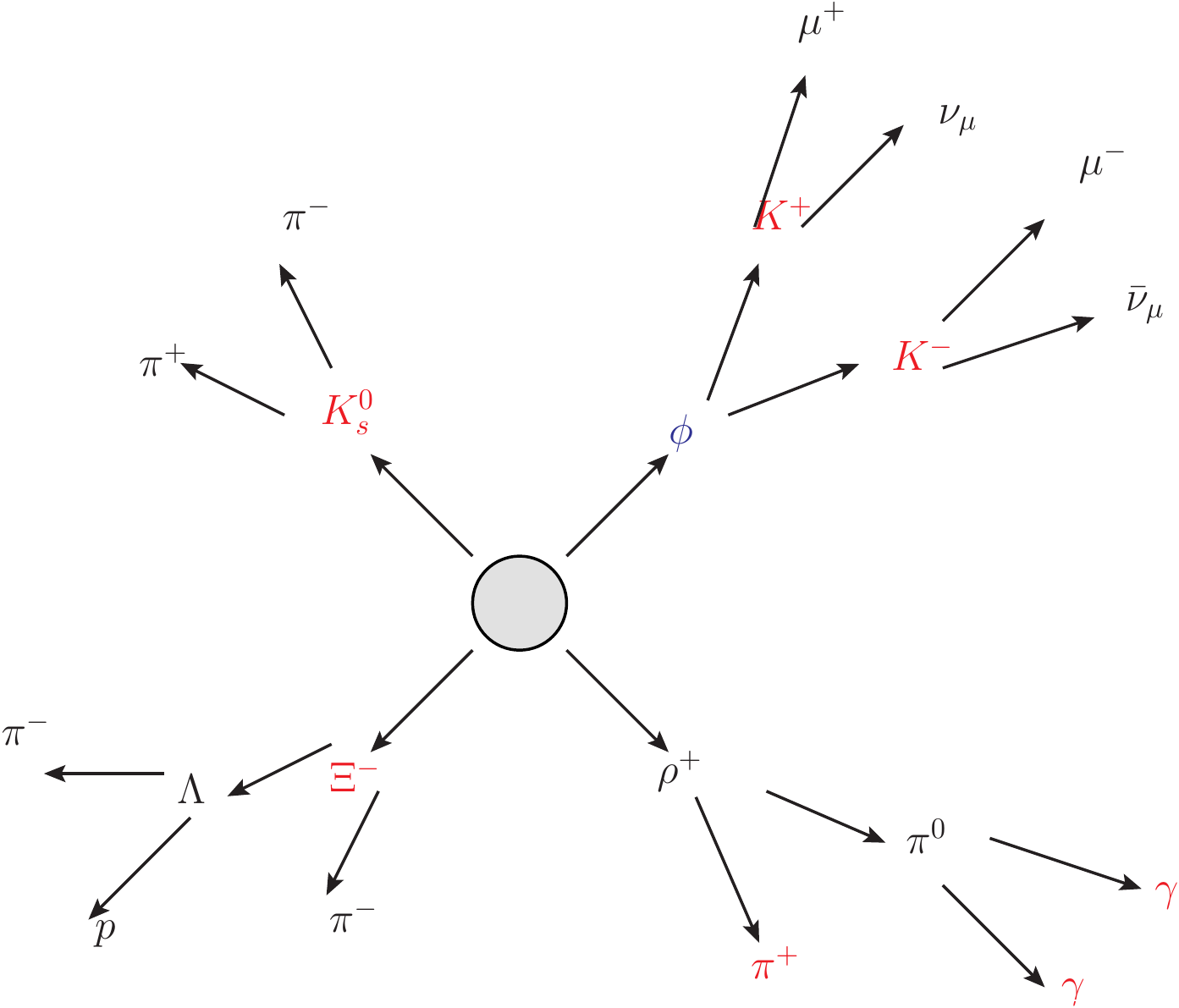}
  \end{center}
  \caption{\label{fig:decay}Visualization of the experimental primary
    particle definition by \alice \cite{ALICE-PUBLIC-2017-005}, with
    primary particles in red, particles not considered primary in
    black and resonances which can possibly be added in addition to
    primary particles in blue.}
\end{figure}

The full event record of the \mc{event generator} cannot be compared
directly to experimental results as these are often corrected to a
certain definition of primary particles, illustrated in
Figure~\ref{fig:decay}.  In the context of the heavy-ion developments,
pre-defined \texttt{Projections} have been developed for the
primary-particle definitions of various experiments. As an example,
the definition by the ALICE experiment \cite{ALICE-PUBLIC-2017-005}
has been implemented in the projection
\texttt{ALICE::PrimaryParticles}, where a particle is considered
primary if it:
\begin{enumerate}
\item has a proper lifetime of $\tau \geq 1\,\mathrm{cm}/c$,
  \emph{and}
\item is \emph{either} produced directly in the interaction;
\item \emph{or} through the decay of a particle with mean proper
  lifetime $\tau < 1\,\mathrm{cm}/c$.
\end{enumerate}

This provides a reusable selection of primary particles for the analysis
of a given experiment. In addition, this approach allows one to clearly
identify cases where a requirement has been relaxed for a specific
analysis. For example, in Ref.~\cite{Acharya:2018orn}, the $\phi$ as
well as its kaon decay daughters have been considered primary.

\subsection{Support for multiple runs through reentrant finalize}
\label{sec:reentrant}

In order to produce statistically significant results with \rivet
analyses, it is often desirable to parallelize the \mc generation step
over multiple computational units, and merge the individual results
into a whole.  If the final result includes anything but simple
objects such as distributions represented as histograms or mean
values, a simple merging will not yield the correct result if the
\texttt{finalize} method (see Section~\ref{sec:technical}) has already
been executed.  Furthermore, a large number of heavy-ion analyses
requires a cross-system merging, to construct, for example, the
nuclear-modification factors (see Section~\ref{sec:example}), which
are ratios of distributions obtained in \AA to the same distribution
in \pp, scaled by an appropriate factor.

For this version of \rivet, both of these requirements have been
addressed.  This solution is called re-entrant finalize, and is
implemented in the script \texttt{rivet-merge}.  The procedure for
re-entrant finalization is as follows:
\begin{enumerate}
\item Before \texttt{finalize}, all output objects are saved in raw
  (un-normalized) form\footnote{The raw (un-normalized) histograms are
    prefaced with the string \texttt{RAW} in the output file.}.
\item Finalize is run as always, and both finalized and raw objects
  are saved in the normal output. Note that raw objects are written
  \textit{before} finalize is executed, and any modifications done in
  finalize will thus not affect the raw objects.
\item By running \texttt{rivet-merge} on one or more produced output
  files, all used analyses are loaded and all raw objects will be
  initialized with summed content of all runs to be merged, and
  \texttt{finalize} will be run again.
\end{enumerate}

For this procedure to work, it is important that all information
needed for finalizing the analysis is properly booked in the analysis
setup. This implies that \emph{only} \yoda objects booked in the
analysis \texttt{init} method can be used in the analysis'
\texttt{finalize}, thus \emph{excluding} plain \Cxx~member objects
(e.g., \texttt{int} or \texttt{std::string}) as these cannot be made
persistent by the framework.

It is worth mentioning that \texttt{rivet-merge} can be run in two
different modes. If the files to be merged contain different
processes (e.g.\ one with \pp events and another one with \AA events for
an $\RAA$ analysis) the program is run in default mode. If, on the
other hand, the files contain the result from several identical runs,
the weight handling is different and the flag \texttt{-e} or
\texttt{----equiv} should be given to indicate this.

\subsection{Analysis options}
\label{sec:analysis-options}

Normally, the implementation of a \rivet analysis is done such that
the user is left with no possibility to influence or modify analysis
logic. In some cases, for example when selecting centrality (see
Section~\ref{sec:centrality}), this dogma would severely restrict the
usability of the framework: A specific combination of analysis and
\mc{event generator} may require a user to inform the analysis about
the nature of the generated events. For that purpose, the possibility
to implement \textit{analysis options} has been implemented.  The
value of an analysis option is specified for each individual analysis
at run-time, and is then used throughout the analysis pass.  Analysis
options can have any type that can be constructed from a string.

In practice, the options are
given as one or more suffixes to the analysis name, as in
\begin{quote}
  \def\meta#1{$\left\langle\text{\itshape #1}\right\rangle$}
  {\ttfamily
    rivet -a \meta{analysis name}:\meta{option 1}=\meta{value
      1}:\meta{option 2}=\meta{value 2}}
\end{quote}
%\begin{verbatim}
%rivet -a AnalysisName:Opt1=Val1:Opt2=Val2
%\end{verbatim}
and the corresponding values for a given option is then available in
the analysis class by calling the
\verb|std::string getOption(std::string)| function. It is possible to
load the same analysis several times, with different option values,
and each instance will then be executed with the corresponding
optional values. If analysis options are given to an analysis, the
output histogram title(s) will be postfixed with a string \verb|[Opt=Val]|
to allow the user to distinguish between histograms generated with
different settings.

\subsection{Event mixing}
\label{sec:event-mixing}
The technique of event mixing is often used when studying correlations
of two (or more) particles.  By mixing data from distinct events, one
can correct for acceptance and efficiency effects in a data-driven fashion. Data
which is compared by \rivet should already be efficiency
corrected. However, the effect of limited acceptance on observables needs to be
estimated and is dependent on the distributions of particles and
therefore on the \mc{generator} used. A simple example is a
rectangular acceptance cut (e.g. $|\eta| < 1$). Sampling two particles
from a uniform distribution in this region and calculating their
pseudorapidity difference, leads to a triangular shape, which has no
relation to the underlying physics. While this particular case can be
analytically calculated, the distribution has a non-trivial shape as
soon as particle production as a function of $\eta$ is not uniform.
Nature, and all state-of-the-art MC generators, falls into the latter category.

The implementation of event mixing in \rivet is based on the usual
strategy at \lhc experiments~\cite{CMS:2012qk}. In this approach the
studied observable is not only built from pairs of particles from a
given event, but also from pairs where the particles are from
different events. In these mixed events all physical correlations are
presumably not present while those originating from the detector
effects remain and are thus estimated in a data-driven way.

One way to illustrate event-mixing analyses is to consider some
$n$-particle correlation measurement $\Delta$
\begin{equationno}
  \Delta(E) = \text{Correlation of $n$ particles in $E$}\quad.
\end{equationno}
The \emph{signal} is then defined as the event-averaged measurement
\begin{equationno}
  S = \langle \Delta(E_i)\rangle_{E}\quad,
\end{equationno}
while the \emph{background} is defined as the average over mixed
events
\begin{equationno}
  B = \langle\Delta(E_i\cap E_j)\rangle_{E}\quad,
\end{equationno}
where $E_i\cap E_j$ is the random mixture of events $E_i$ and $E_j$.  In the
final result
\begin{equationno}
  R = \frac{S}{B}\quad,
\end{equationno}
the random correlations are factored out.

The \rivet implementation allows for booking a projection which
supplies the mixed event sample, by storing events in memory. The
mixed event sample can be divided into bins based on any event
property such that the mixed event sample is taken from events which
are similar to the signal event in question. Common use cases are
centrality and multiplicity, but it is possible to use other
properties, for example event shapes (such as eg. transverse
spherocity), should the need be.

Finally the possibility that the \mc{event generator} supplies
weighted events is considered. If a sample contains weighted events,
the particle content of the mixed sample will be weighted as well,
with particles being selected to the mixed sample with a probability
given by $w_e/\sum_i w_i$, where $w_e$ is the weight of the event
containing the signal, and $w_i$ are the weights off all events in the
ensemble of events that form the mixed event.

\section{Specialized features for heavy-ion collisions}
\label{sec:features}
Heavy-ion analyses commonly make use of concepts often not considered
relevant for high energy particle physics and, thus, has been not available in
\rivet.  In this section two new features motivated by physics
requirements are outlined. In Section~\ref{sec:centrality} the
framework for centrality is described, and in
Section~\ref{sec:generic-framework} the generic framework for flow
measurements is briefly introduced, and the implementation outlined.

\subsection{Centrality framework}
\label{sec:centrality}
The impact parameter ($b$) is the distance between the centers of the
colliding nuclei. Its value determines on average the size and the
transverse shape of the interaction region of a heavy-ion
collision. Observables' dependence on the geometry of the interaction
region, provides insight into the effect of the size and geometry on
the underlying physical mechanisms.

Experimentally, centrality classifies collisions on the basis of some
measurement (or measurements) which is assumed to depend
monotonically\footnote{The monotonic dependence may be over a limited
  range, for example in \emph{zero-degree calorimeters}.}  on the
impact parameter $b$ of a given collision.  For a single measurement
$X$ which is assumed large for small $b$ and small for large $b$, we
can define

\begin{equation}
  \label{eq:n-cent}
  c_{\mrm{exp}} =\frac{1}{\sigma_{\mathrm{vis}}}
  \int_{X}^{\infty}\mathrm{d}X\,
  \frac{\mathrm{d}\sigma_{\mathrm{vis}}}{\mathrm{dX'}}
  = \frac{1}{N}\int_{X}^\infty\mathrm{d}X'
  \frac{\mathrm{d}N}{\mathrm{d}X'}
  \quad,
\end{equation}
where $\mathrm{d}\sigma_{\mathrm{vis}}/\mathrm{d}X$ is the
full \emph{measured} distribution of $X$ over the visible
cross-section $\sigma_{\mathrm{vis}}$ (typically defined in terms of
minimum-bias trigger conditions) of the experiment, and $X$ is the single
collision measurement.  From a \emph{theoretical} point of view, the
centrality of a single collision can be defined as
\begin{equation}
  \label{eq:b-cent}
  c_{b} = \frac{1}{\sigma_{\mrm{inel}}} \int_0^b \mrm{d}b'\,
  \frac{\mrm{d}\sigma_{\mrm{inel}}}{\mrm{d}b'}\quad.
\end{equation}
If $\sigma_{\mathrm{vis}}\approx\sigma_{\mathrm{inel}}$ and $X(b)$ is reasonably well
behaved (smooth, positive, monotonically decreasing, with appropriate limiting behaviour) \cite{Das:2017ned},
$c_b \approx c_{\mathrm{exp}}$.

Different experiments use different observables for $X$. In \rivet,
several experimental definitions are available, corresponding to
different definitions listed in the following. The \alice
collaboration~\cite{Abelev:2013qoq} uses the integrated signal in the
V0 scintillator systems covering $-3.7 < \eta < -1.7$ and
$2.8 < \eta < 5.1$. The \star collaboration~\cite{Abelev:2006cs} uses
the number of charged tracks in the central pseudorapidity interval
spanning $-0.5 < \eta < 0.5$, while \brahms used the charged-particle
multiplicity at mid-rapidity
\cite{Bearden:2004yx,Adamczyk:2003sq}. The \atlas collaboration
\cite{ATLAS:2011ah} uses transverse energy deposited in the forward calorimeters
located at $4.9 < |\eta| < 3.2$.

In central heavy ion collisions $c_{b} = c_{\mrm{exp}}$ to good
precision \cite{Das:2017ned}. In peripheral collisions, or in \pA and
\pp collisions the correspondence is less obvious~\cite{Bierlich:2018xfw}.
Moreover, biases introduced by experimental trigger conditions for measurements of $X$,
are by construction not included in equation~(\ref{eq:b-cent}). When
possible, it should therefore always be preferred to use
equation~(\ref{eq:n-cent}) (with appropriate trigger conditions) for
estimation of centrality in a simulated event, essentially comparing
experimentally obtained $X$ directly to $X$ obtained in a \mc. By
construction, such an approach also includes biases not related to the
detector used, but to the definition of $X$~\cite{Adam:2014qja}.

\begin{figure}
  \includegraphics[width=\twofigw]{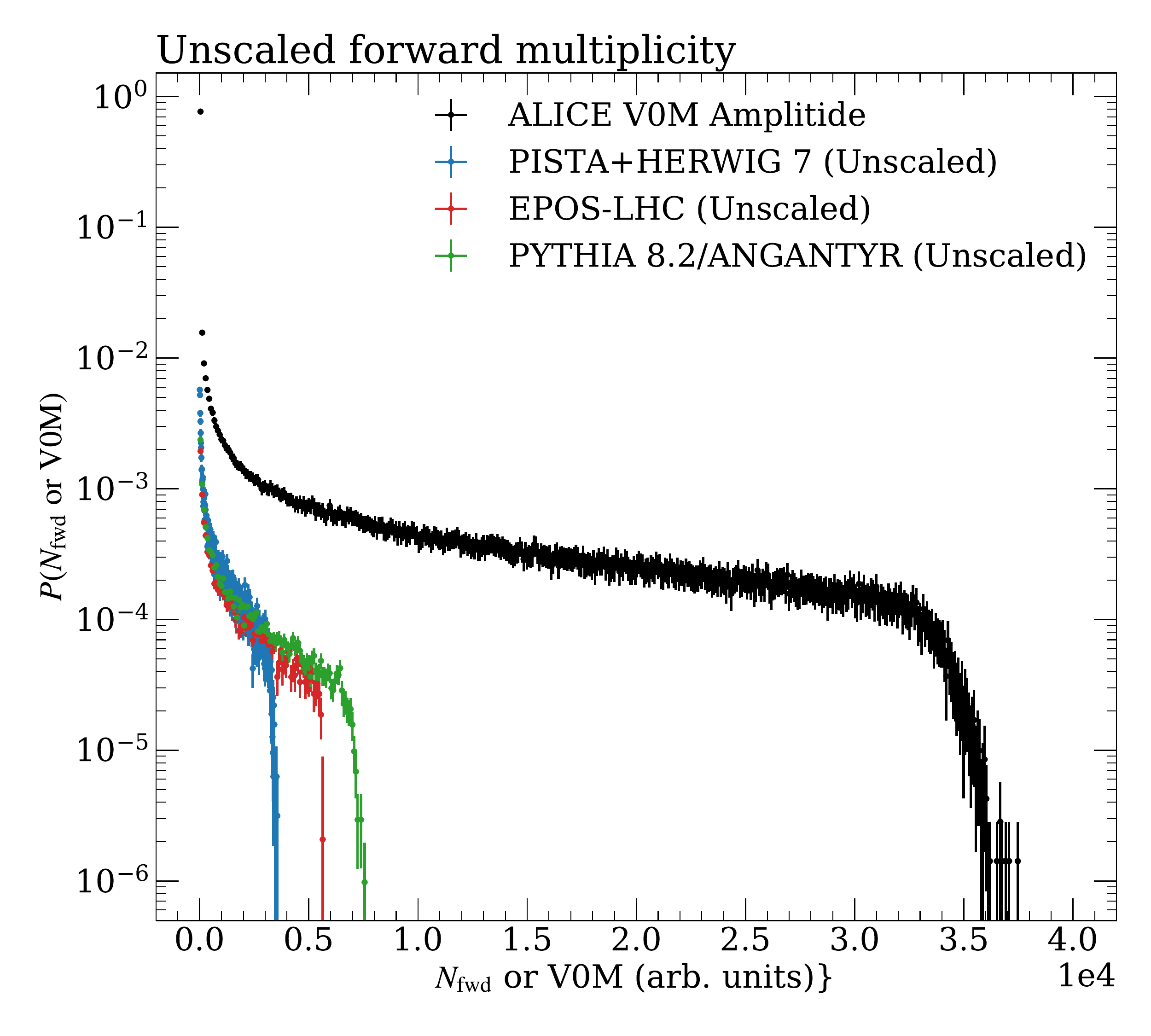}
  \includegraphics[width=\twofigw]{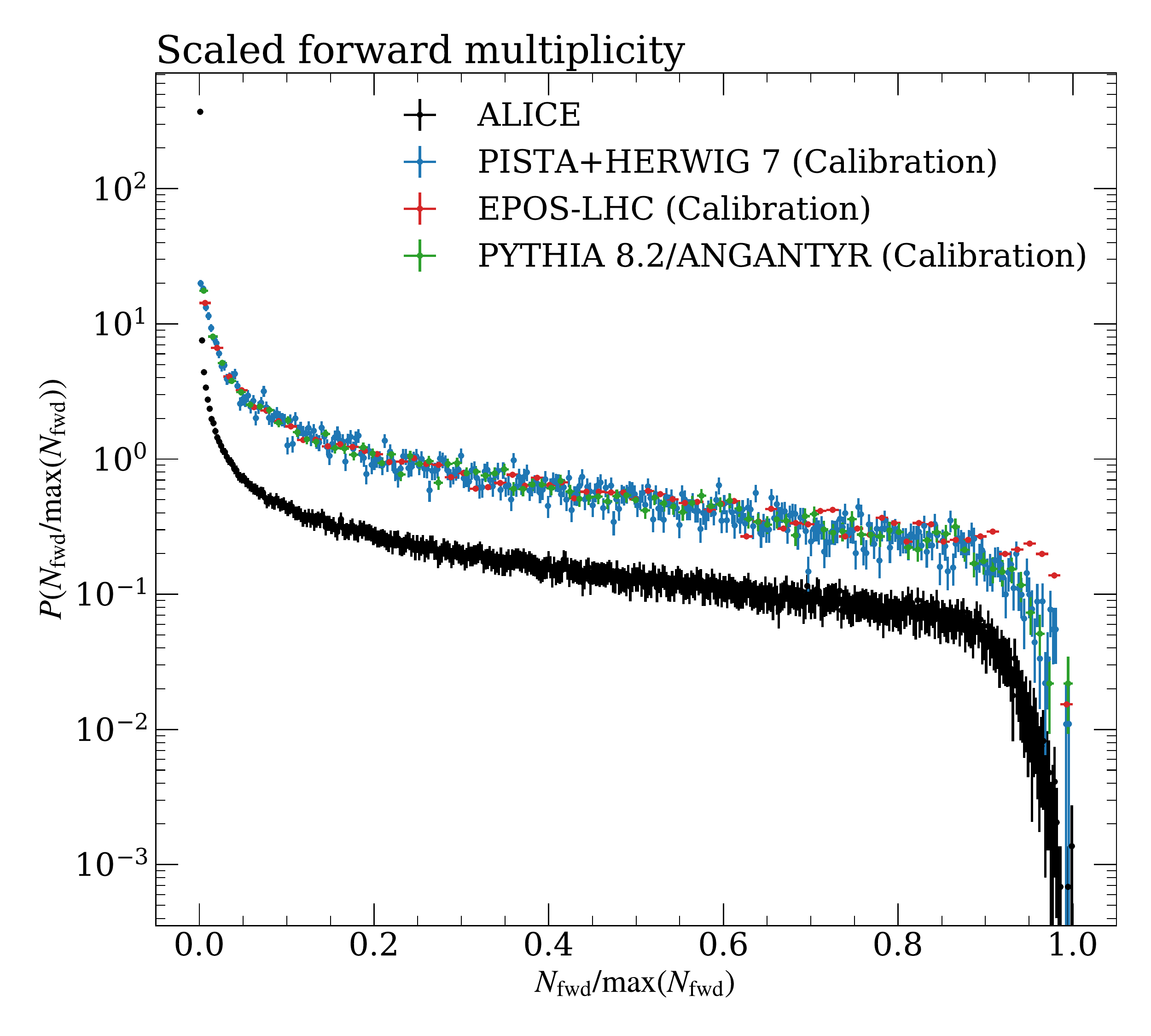}
  \caption{\label{fig:centrality}Comparison of generated centrality
    measure generated by \herwigpista, \eposlhc and \angantyr
    to V0M Amplititude from ALICE. To the left a direct comparison,
    and to the right a rescaled one (see text).}
\end{figure}

In some cases, neither using equation (\ref{eq:b-cent}) nor equation (\ref{eq:n-cent}), is an option. For
example, the event generator \jewel~\cite{Zapp:2013vla} does not
calculate the underlying event, but considers only a jet quenched in a
medium at a certain impact parameter. Other generators, for example
\herwigpista~\cite{Bellm:2018sjt,Bellm:2015jjp} attempt to generate
the forward-going particle multiplicity but do not obtain good
agreement with experimental results.  In other cases still, it is not
possible to compare to the experimentally measured $X$ without taking
into account the detector response. This is for example the case for
\alice, where $X$ is defined as a detector response (affected by
secondary particles), not unfolded to particle level. This latter case
is illustrated in Figure~\ref{fig:centrality}, where three soft
inclusive approaches (\herwigpista, \eposlhc and \angantyr)
are shown together with centrality data from ALICE.  In Figure~\ref{fig:centrality}
(left) the raw detector response is overlaid with
the centrality measure implemented in \rivet, and in Figure~\ref{fig:centrality}
(right), a rescaled, normalized version of the
centrality measure is shown, comparing to the three MCs.  It is clear
that direct comparison of MC to experimentally obtained measure (as in
Figure~\ref{fig:centrality} (left)) has no physical meaning at all,
since the scale of the compared quantities are different, and in order
to refrain from potential errors, this should be avoided in analyses.
In such cases, a better choice is to use equation~\ref{eq:n-cent} on
\mc and compare the extracted $c_{\mrm{exp}}$.  Note, however, that it
is important then to implement the appropriate criteria for
$\sigma_{\mathrm{vis}}$ in an equivalent fashion to the experiment
compared to.  The large excess of experimental measurements at low $X$
are due to contamination from electromagnetic interactions.  However,
as the analyses often ignore this region, it is less of a problem in
direct comparisons to \mc results.  Remnant contributions from such
electromagnetic interactions in more central events, are accounted for
by appropriate systematic uncertainties on the experimental results.

In consequence, the slicing from data cannot be naively applied to
\mc, making the calibration step necessary. In total, four possibilities
for centrality estimation can be used:

\begin{enumerate}
\item Direct comparison to, and binning in, measured $X$.  This, to
  some extend, circumvents the notion of centrality altogether, but
  requires that the simulated and measured $X$ are roughly equally
  distributed. In \rivet, this is selected by the analysis option
  \texttt{cent=REF}.
\item Comparison to a user-implemented $X$, which should be a particle
  level, final state observable to get \emph{generator calibrated}
  centrality.  As noted above, it is important to impose the same
  requirements as used by the experiment to determine
  $\sigma_{\mathrm{vis}}$\footnote{For example, the projection
    \texttt{ALICE::V0AndTrigger} selects the \alice
    $\sigma_{\mathrm{vis}}$.}. In \rivet, this is selected by the
  analysis option \texttt{cent=GEN}.
\item Centrality binning of \mc according to
  equation~(\ref{eq:b-cent}) --- $b$ centrality. In \rivet, this is
  selected by the analysis option \texttt{cent=IMP}.
\item Provide a centrality percentile number in the \hepmc~heavy-ion
  record (available from \hepmc~3.0). In \rivet, this is selected by
  the analysis option \texttt{cent=RAW}.
\end{enumerate}

Possibilities 2 and 3 require the user to perform a calibration run,
to generate the distribution of either $X$ or $b$ and then evaluate
the integrals in equations~(\ref{eq:b-cent}) or (\ref{eq:n-cent}). The
result of the calibration pass can then be used in subsequent passes
with the analysis calculating the desired centrality-dependent
observable. Concrete examples of how to perform such tasks using
\rivet, are given in Section~\ref{sec:example}.

\subsection{Flow measurements}
\label{sec:generic-framework}
In heavy-ion collisions, the study of azimuthal anisotropy of particle
production is used to understand the transport properties of the
produced medium, as well as the geometry of the initial
state~\cite{Ollitrault:1992bk}.  This is quantified in flow
coefficients $v_n$'s, obtained by a Fourier expansion of the particle
yield with respect to the reaction plane $\Psi_n$
\begin{equation}
  \label{eq:flowdefinition}
  E\frac{\mrm{d}^3N}{\mrm{d}^3p} =
  \frac{1}{2\pi}\frac{\mrm{d}^2N}{\pT \mrm{d}\pT \mrm{d}y}
  \left(1 + 2 \sum_{n=1}^\infty v_n \cos([n(\phi - \Psi_n)])\right)\quad,
\end{equation}
where $E$ is the energy of the particle, $\pT$ is the transverse
momentum, $\phi$ is the azimuthal angle and $y$ is the rapidity. As indicated, the coefficients are
in general dependent on $y$ and $\pT$.  The present \rivet
implementation allows one to perform a simple calculation of coefficients
differentially with respect to $\pT$, as well as integrated coefficents.

Since the reaction plane cannot be determined experimentally, flow
coefficients are estimated from two- or multi-particle correlations.
For example, using two-particle correlations:
\begin{equation}
  \label{eq:naive}
  \langle v^2_n \rangle \approx
  \langle \cos(n(\phi_1 - \phi_2)))\rangle =
  \langle \exp(in(\phi_1 - \phi_2))\rangle\quad,
\end{equation}
where $\phi_1 = \phi_2$ is excluded from the average. A more advanced
technique -- with several benefits, which are introduced in the
following -- calculates multi-particle cumulants from $Q$-vectors
\cite{Bilandzic:2010jr} ($Q_{n} = \sum_{k=1}^M w_k\exp(in\phi_k)$ for
an event with $M$ particles)\footnote{In most experimental treatments,
  the $Q$-vectors, and following the correlators are introduced with a
  particle weight, to correct for detector inefficiencies. Since
  \rivet compares \mc, which suffers no detector inefficiencies, to
  corrected data, the weight is left out in this summary.  The
  implementation does, however, allow for weights.}, which are combined
into $m$-particle correlators denoted
$\langle m \rangle_{n_1,n_2,...,n_m}$. Expressing $m$-particle
correlators in terms of $Q$-vectors follows the Generic
Framework~\cite{Bilandzic:2013kga}, where the formula for any
$m$-particle correlator is written as:
\begin{align}
  \label{eq:gf}
  \langle m \rangle_{n_1,n_2,...n_m}
  &= \frac{N\langle m \rangle_{n_1,n_2,\ldots,n_m}}{
    N \langle m \rangle_{0,0,\ldots,0}}\quad\text{with} \\
  N\langle m \rangle_{n_1,n_2,\ldots,n_m}
  &= \sum_{\substack{k_1, k_2,\ldots,k_m = 1 \\ k_1 \neq k_2 \neq
  \ldots k_m}}^M
  \exp(i(n_1\phi_{k_1} + n_2\phi_{k_2} + \cdots + n_m\phi_{k_m}))\quad.
\end{align}
This expression allows one to easily calculate any correlator, since
$N\langle m \rangle_{n_1,n_2,...,n_m}$ can be calculated from
$Q$-vectors, with the concrete expressions obtained using a recursion
relation~\cite{Bilandzic:2013kga}, for any values of $m$ and $n_i$. As
an example, for $m = 2$, one obtains:
\begin{equation}
  \label{eq:Nm}
  N\langle 2 \rangle_{n_1,n_2} = Q_{n_1}Q_{n_2} - Q_{n_1 + n_2}\quad.
\end{equation}
For the special case of $n_1 = n_2 = n$, the two-particle correlator
directly becomes:
\begin{equationno}
  \langle 2 \rangle_{n} = \frac{|Q_n|^2 - M}{M(M-1)}\quad.
\end{equationno}
Evaluation of the product in the numerator
$|Q_n|^2 = \sum_{k,l} \exp(in(\phi_k - \phi_l))$ reveals that the
correlator reduces to equation~(\ref{eq:naive}).  However, when evaluating
using $Q$-vectors, the process requires only a single pass over data, as
opposed to nested passes.  Without the use of $Q$-vectors, the
complexity of an $m$-particle correlator is obviously
$O\left(M^m\right)$ while the above approach is at most
$O\left(M\log M\right)$\footnote{Note, however, that the evaluation of
  the $Q$-vectors themselves are generally $O\left(mM\right)$, and
  considerable computing time is spent building the event
  $Q$-vectors.}.  In a heavy-ion event where $M$ can be 10\,000
tracks, this is absolutely crucial.  From the correlators, a large
number of flow observables can be defined. In most analyses, cumulants
$c_n\{m\}$ (where curly braces represent the order of the $m$-particle
correlation) are calculated from correlators with
$n_1 = n_2 = ... = n_m = n$. The most common use cases are implemented
along with uncertainty estimation (see below):

\def\dblavg#1{\langle\langle #1\rangle\rangle}
\begin{align}
  \label{eq:c22}
  c_n\{2\} =
  & \dblavg{2}_{n,-n} \nonumber\\
  c_n\{4\} =
  & \dblavg{4}_{n,n,-n,-n}- 2\dblavg{2}^2_{n,-n}\nonumber\\
  c_n\{6\} =
  & \dblavg{6}_{n,n,n,-n,-n,-n}
    -9 \dblavg{2}_{n,-n}\dblavg{4}_{n,n,-n,-n}
    + 12 \dblavg{2}^3_{n,-n} \nonumber\\
  c_n\{8\} =
  & \dblavg{8 }_{n,n,n,n,-n,-n,-n}
    - 16\dblavg{6}_{n,n,n,-n,-n,-n}\dblavg{2}_{n,-n}
    - 18\dblavg{4}^2_{n,n,-n,-n} \nonumber\\
  &{}+ 144 \dblavg{4}_{n,n,-n,-n} \dblavg{2}^2_{n,-n}
    - 144\dblavg{2}^4_{n,-n}\quad.
\end{align}
The double averages denote an all-event average of event
averages. Even higher-order cumulants, or other combinations of
correlators, such as Symmetric Cumulants~\cite{ALICE:2016kpq} can be
easily added by the user, as the framework allows the user to fully 
access the correlators. The cumulants are finally transformed into flow
coefficients
\begin{equationno}
  v_n\{m\} = \left (k_{m/2} c_n\{m\} \right)^\frac{1}{m}\quad,
\end{equationno}
where the coefficient $k_m$ is written here for convenience for
$m < 4$:
\begin{equationno}
  k_m = \left\{1, -1, \frac{1}{4}, -\frac{1}{33}\right\}\quad.
\end{equationno}

Differential flow observables are constructed using an additional
correlator, limited to a certain phase space. In the case of the
implementation in \rivet, this is restricted to observables
differential in $\pT$. The differential correlator can be constructed
in a similar fashion~\cite{Bilandzic:2013kga} as the integrated ones
in equation~(\ref{eq:gf}), but the transformation to differential flow
coefficients differs. Denoting the differential $m$-particle
correlator $\langle m' \rangle$, the differential two-particle
cumulant and the corresponding differential flow coefficient are given
by:
\begin{equationno}
  d_n\{2\} = \dblavg{2' }
  \quad\text{and}\quad
  v'_n\{2\} = \frac{d_n\{2\}}{\sqrt{c_n\{2\}}}\quad,
\end{equationno}
where $c_n\{2\}$ is a reference cumulant calculated for the full phase
space as above. In the same way, the differential four-particle
cumulant and the corresponding differential flow coeffiecient are:
\begin{equationno}
  d_n\{4\} = \dblavg{4'} - 2\dblavg{2} \dblavg{2' }
  \quad\text{and}\quad
  v'_n\{4\} = -\frac{d_n\{4\}}{(-c_n\{4\})^{3/4}}\quad.
\end{equationno}

The statistical uncertainty on cumulants and flow coefficients is
calculated by a simple variance method, often denoted as
bootstrap~\cite{LWasserman} method in experimental literature. Instead
of propagating the sampling uncertainty on the individual correlators
through the (long) expressions for cumulants and flow coefficients,
the calculation is internally split up into sub-samples\footnote{An
  alternative to the experimental practise of using ``bootstrap''
  could be replaced by error propagation as outlined in
  ref.~\cite{Bilandzic:2013kga}, and its reference implementation. In
  \rivet, the choice is taken to stay as close to the performed
  experimental analyses as possible.}. The default behaviour is to
display the square root of the sample variance as statistical
uncertainty, but the user can choose to use the envelope instead.

\subsubsection{Suppressing non-flow with an $\eta$-gap}
The approximate equality in equation~(\ref{eq:naive}) becomes a full
equality in the absence of non-flow effects, which are correlations
arising from other sources than typically attributed to collective
expansion~\cite{Voloshin:2008dg}.  Non-flow effects can arise from
several sources, such as jet production or resonance decay, and are
few-particle correlations as opposed to flow which is correlation of
all particles to a common symmetry plane. The idea behind the
$\eta$-gap or sub-event method is to suppress sources of non-flow
which are local in (pseudo)rapidity.  The non-flow is then suppressed
by requiring that particle measurements contributing to a correlator
\emph{must} come from different parts of the detector acceptance,
usually separated in pseudorapidity.  Using equation~(\ref{eq:naive}),
an $\eta$-gap can be implemented simply by requiring that the two
particles come from two different sub-events, here labeled $A$ and $B$:
\begin{equationno}
  \langle v^2_n \rangle \approx \langle \exp(in(\phi^A_1 -
  \phi^B_2))\rangle\quad.
\end{equationno}
Translating to $Q$-vectors, it is first noted that $\phi_1 = \phi_2$
does not need to be excluded from the average in this case, as it is
excluded by construction. Equation~(\ref{eq:Nm}) reduces to
$N\langle 2 \rangle_{n_1,n_2} = Q^A_{n_1}Q^A_{n_2}$, where
superscripts $A$ and $B$ indicates that $Q$-vectors are calculated in
that specific sub-event only, as
$Q^X_n = \sum^{M_X}_{k=1} \exp(in\phi^X_k)$. This procedure is
generalized to multi-particle correlations, by removing terms from the
correlator where cross-terms do not share a common sub-event.

In \rivet, correlators with two sub-events are implemented, with the
possibility of calculating both integrated and differential flow.

\begin{figure}
  \includegraphics[width=\twofigw]{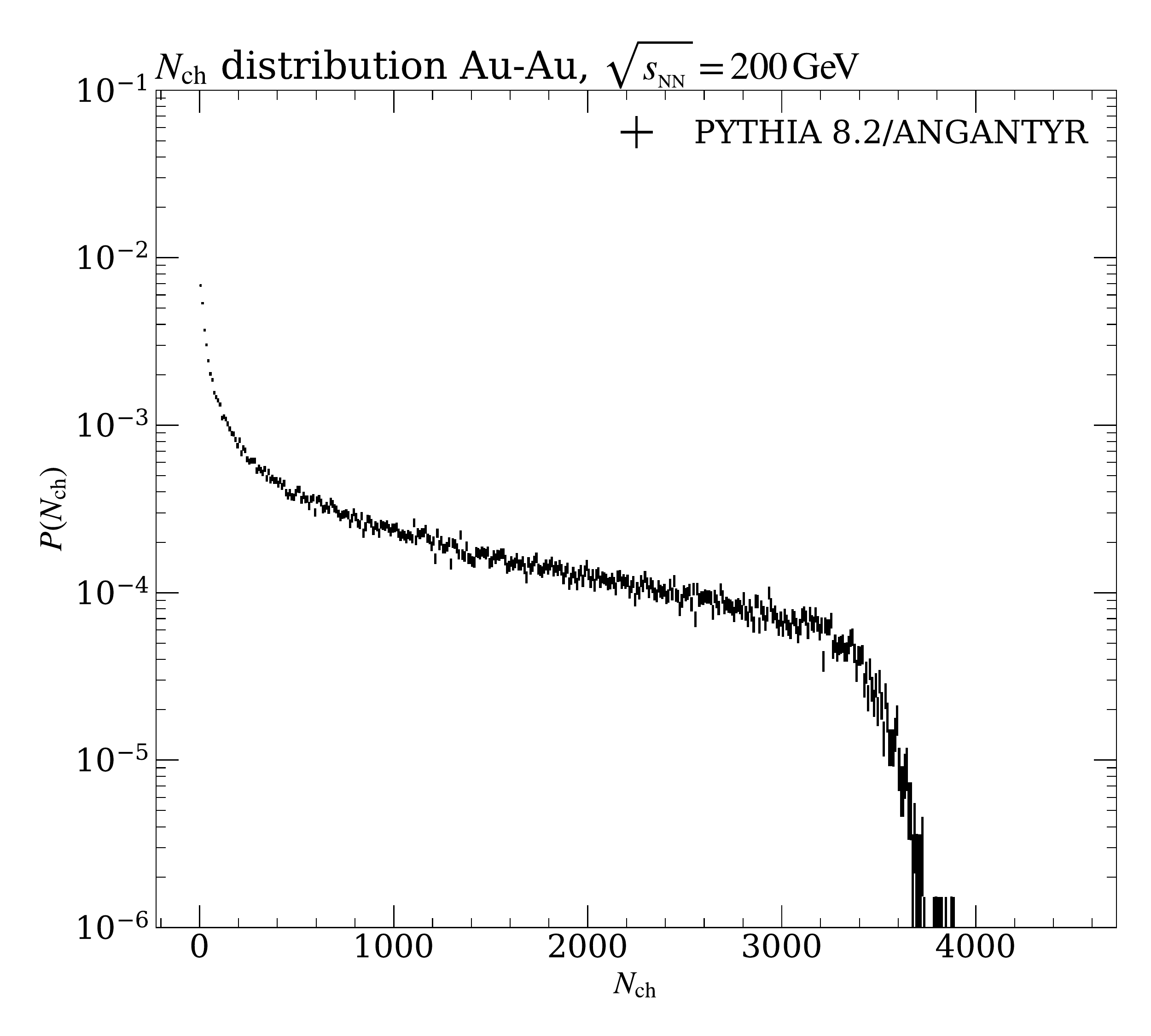}
  \includegraphics[width=\twofigw]{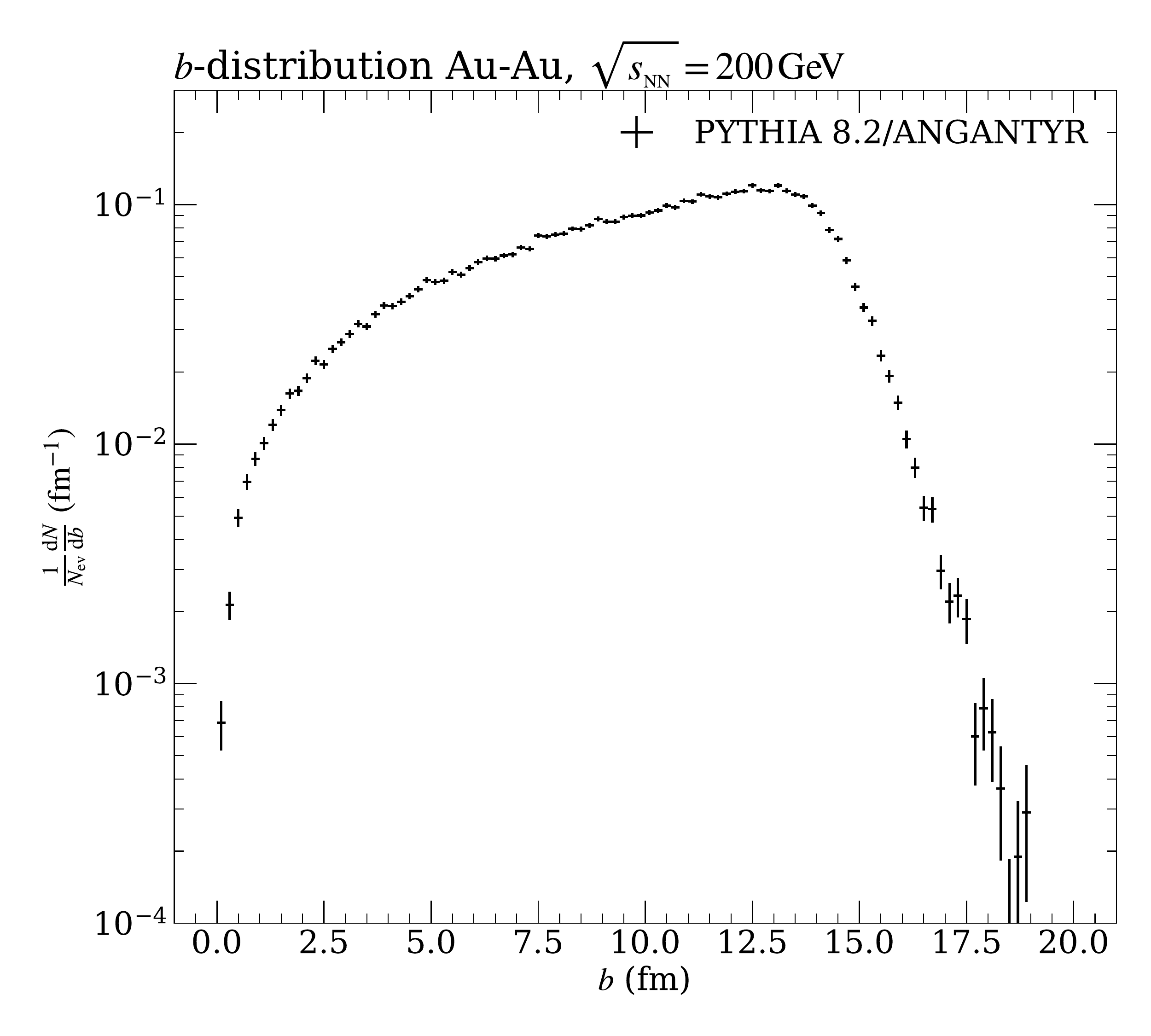}
  \caption{\label{fig:cent} Centrality observables from the analyses
    \texttt{BRAHMS\_2004\_AUAUCentrality} with \angantyr
    model \cite{Bierlich:2018xfw} displaying the generated calibration
    curve (left) and the impact parameter calibration curve (right).}
\end{figure}

\section{Examples of implemented heavy-ion analyses}
\label{sec:example}

In this section, some of the implemented \rivet analyses using new
features are presented, along with detailed instructions for a user to
run the analyses and plot the resulting figures.

The \emph{centrality framework} introduced in Section~\ref{sec:centrality}
is used by many analyses, each one implementing
the centrality estimator of the respective experiment. In these cases
the analysis implementation relies on an analysis for centrality
calibration, called e.g., \texttt{ALICE\_2015\_PBPBCentrality},
\texttt{BRAHMS\_2004\_AUAUCentrality}, and
\texttt{STAR\_BES\_CALIB}. These can be reused by many analyses
sharing the same centrality estimator.

Executing the centrality calibration analysis fills histograms for
each defined centrality estimator, for example the impact parameter
and the number of charged particles or their energy in the rapidity
region where the relevant detector is situated. Figure~\ref{fig:cent}
exemplifies the calibration for \brahms, for which an unfolded measure
does not exist, and the options 2 or 3 (\texttt{GEN} and \texttt{IMP})
must be used.

In order to use the generated calibration histograms in the actual
analyses, they must be pre-loaded when running the analysis code.  The
analysis will then calculate the percentile of the calibration curve
for the value determined in the newly generated event and, thus, allow
the user to retrieve the centrality for the current event.

\begin{figure}
  \includegraphics[width=\threefigw]{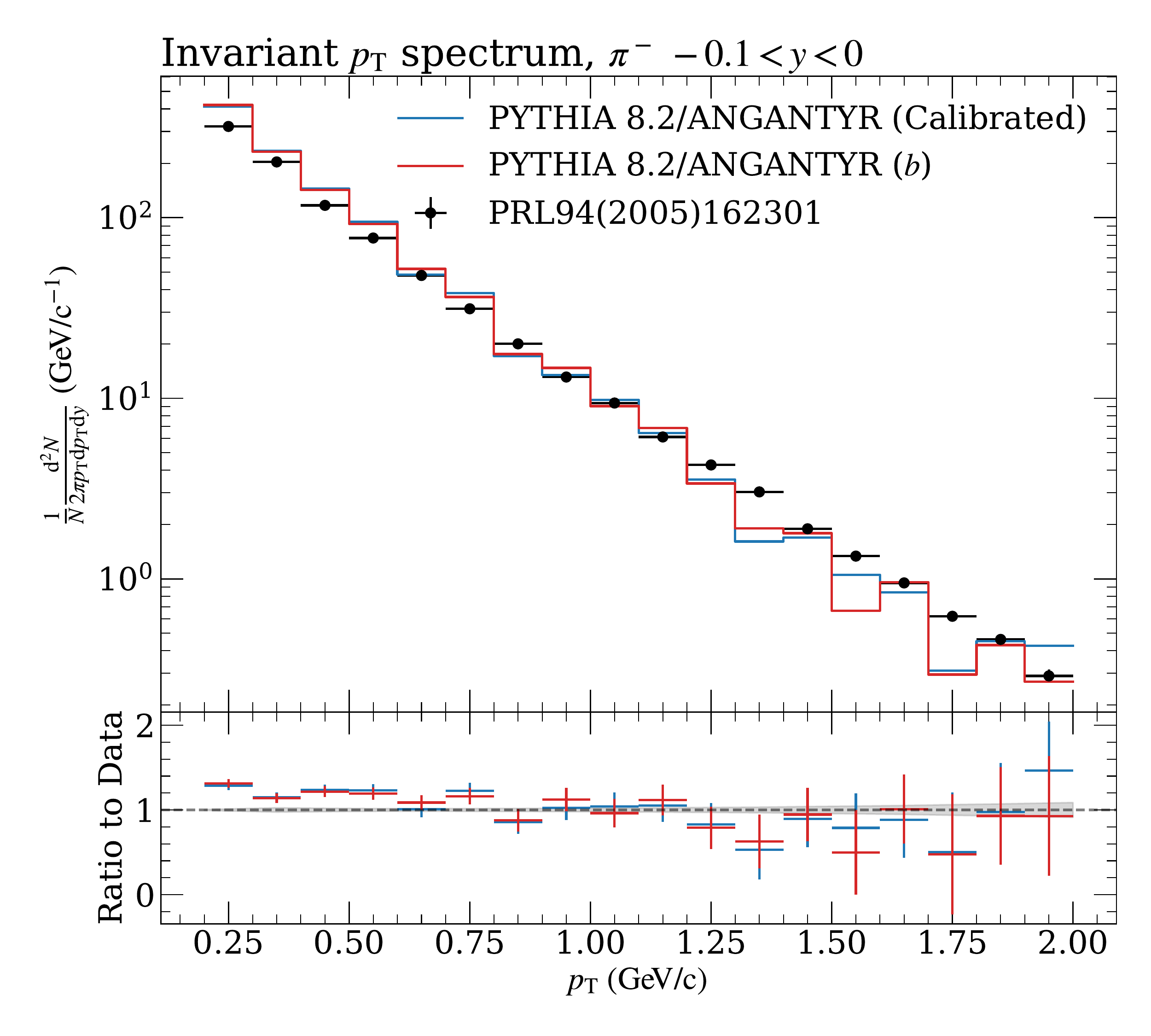}
  \includegraphics[width=\threefigw]{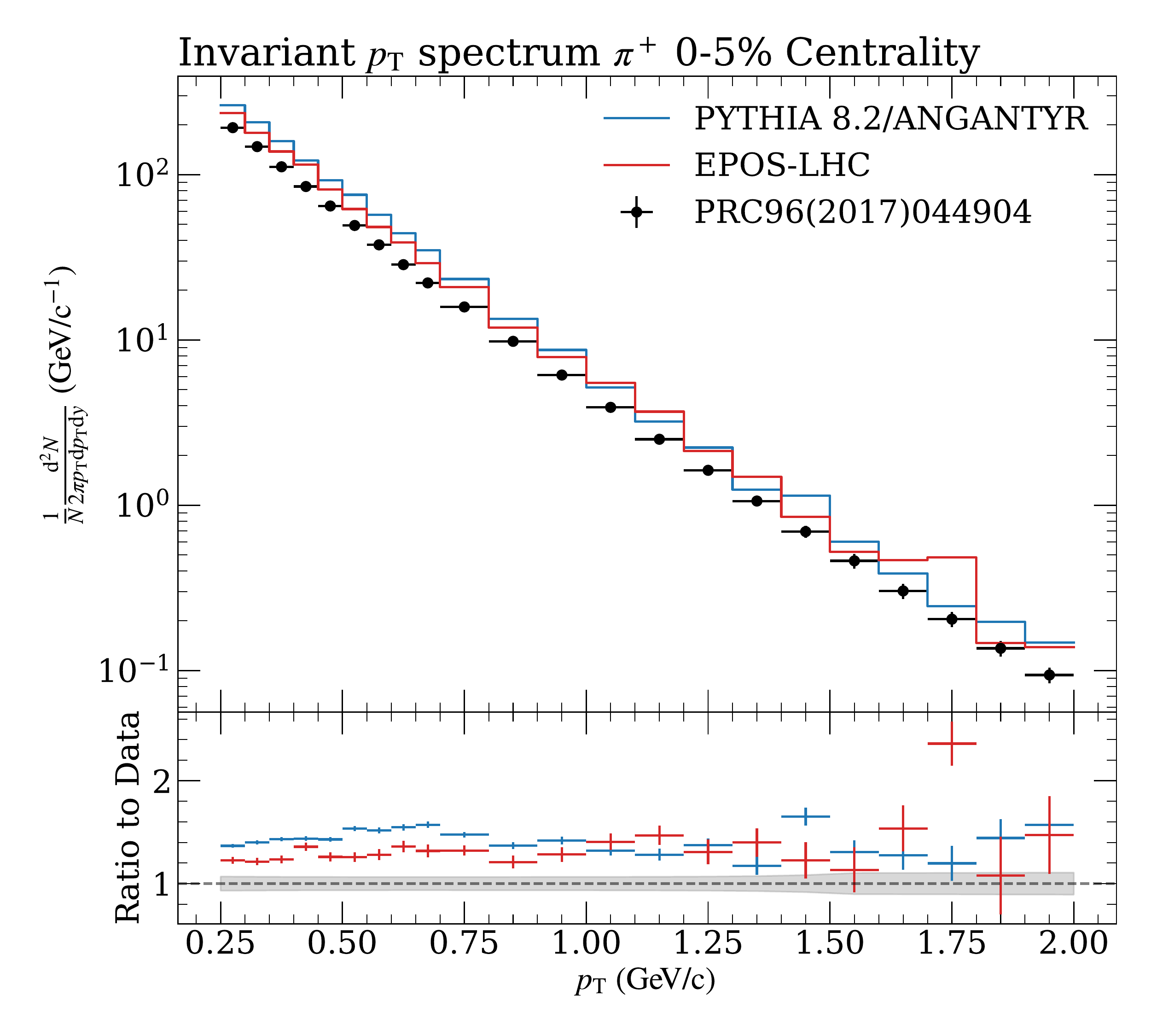}
  \includegraphics[width=\threefigw]{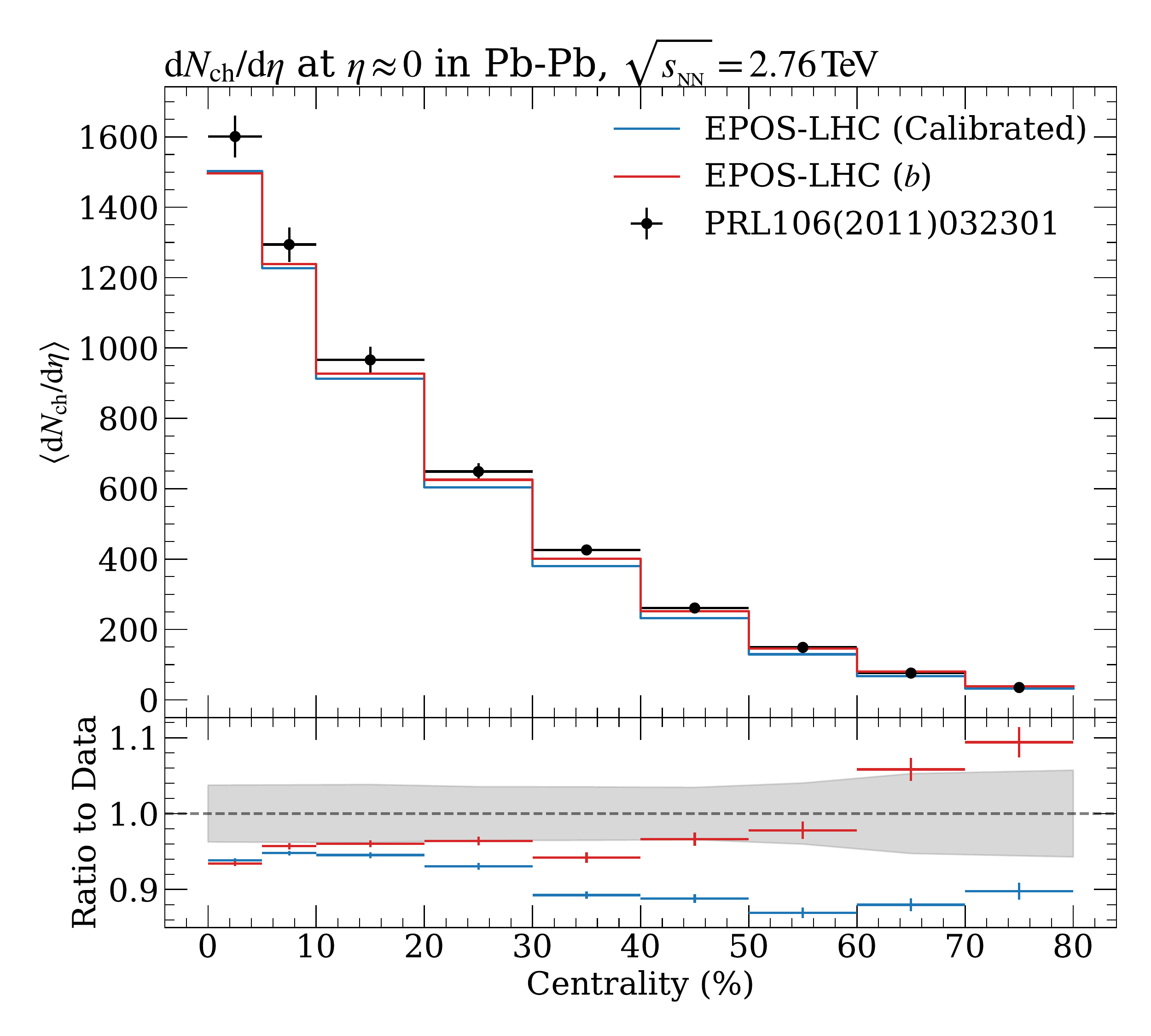}
  \caption{\label{fig:centrality-res} Results from the analyses
    \texttt{BRAHMS\_2004\_I647076} (left; invariant $\pT$ of
    $\pi^-$ in $0 < y < 0.1$), \texttt{STAR\_2017\_I151059} (middle;
    invariant $\pT$ for $K^+$ in $|y| < 0.1$) and
    \texttt{ALICE\_2010\_I880049} (right;
    $\mathrm{d}N_{\mathrm{ch}}/\mathrm{d}\eta$ vs. centrality)
    utilizing centrality calibration.}
\end{figure}

Figure~\ref{fig:centrality-res} presents results from \brahms, \star
and \alice, using separate centrality calibrations.  Comparison to
\angantyr and \eposlhc is performed. In the considered analysis by
\brahms (\texttt{BRAHMS\_2004\_I647076}~\cite{Bearden:2004yx}), only
the $0-5\%$ centrality bin is used, meaning that all other events are
vetoed by the analysis code. The centrality definition is selected at
runtime using, in this case, either the analysis option (see
Section~\ref{sec:analysis-options}) \texttt{GEN} (for generator calibrated) or
\texttt{IMP} (for impact parameter). In the \star analysis
\texttt{STAR\_2017\_I1510593}~\cite{Adamczyk:2017iwn}, several
centrality bins are analysed. The \alice analysis
\texttt{ALICE\_2010\_I880049}~\cite{Aamodt:2010cz} calibrates using
the centrality analyses presented in Section~\ref{sec:centrality}
(cf.~Figure~\ref{fig:centrality}).

To reproduce those results based on a \hepmc file containing \AA
collisions generated by the considered generators, the following steps
need to be followed for the example of the \brahms analysis:
\begin{enumerate}
\item Run the calibration analysis
\begin{verbatim}
rivet events.hepmc -a BRAHMS_2004_AUAUCentrality -o calibration.yoda
\end{verbatim}
\item Plot the calibration results from Figure~\ref{fig:cent}
\begin{verbatim}
rivet-mkhtml calibration.yoda
\end{verbatim}
\item Run the analysis with the calibration file pre-loaded, for both
  centrality selections:
\begin{verbatim}
rivet events.hepmc -p calibration.yoda  \
      -a BRAHMS_2004_I647076:cent=IMP \
      -a BRAHMS_2004_I647076:cent=GEN -o result
\end{verbatim}
\item Plot the analysis result\footnote{Note that the figures in this
    paper are produced using Python/Matplotlib, to allow for rebinning
    and rescaling of results for illustrative purposes. Equivalent
    figures (apart from those requiring such steps) can be produced
    using \texttt{rivet-mkhtml}.} from Figure~\ref{fig:centrality-res}
\begin{verbatim}
rivet-mkhtml result.yoda
\end{verbatim}
\end{enumerate}
Note that in this case (and in fact in all currently implemented heavy ion
analyses), the calibration and analysis steps are carried out on the same
event sample. This would not be the case for a signal analysis considering
a signal process (eg. $Z$-production). Here the calibration step would be
carried out on a minimum bias sample, and the signal step on a signal sample.

The real power of \rivet analyses is, however, not to compare a single
generator at a time, but the ability to compare several generators to
the same experimental data at once. This use case is shown in Figure~\ref{fig:centrality-res}
(middle), where both \eposlhc and
\angantyr are compared to \AuAu~data at $\sNN{27.6}{\GeV}$
obtained by \star~\cite{Adamczyk:2017iwn}. In this case, the
calibration must be run for both generators, as the output will in
general be different. The shown figure can thus be obtained by
slightly extending the workflow outlined above:
\begin{verbatim}
rivet epos-events.hepmc -a STAR_BES_CALIB -o epos-calib
rivet pythia-events.hepmc -a STAR_BES_CALIB -o pythia-calib
rivet epos-events.hepmc -p epos-calib.yoda -a STAR_2017_I1510593 \
  -o epos-result
rivet pythia-events.hepmc -p pythia-calib.yoda -a STAR_2017_I1510593 \ 
  -o pythia-result
rivet-mkhtml epos-result.yoda pythia-result.yoda
\end{verbatim}
Note from the last line that comparison plots can be produced directly
by the \texttt{rivet-mkhtml} script. Such comparisons are performed
using large scale volountary computing resources as part of the
\mcplots project~\cite{Karneyeu:2013aha}. \mcplots is currently being
updated and extented to include heavy-ion results and \mc by members of
the ALICE collaboration\footnote{The
  original \mcplots can be found at \url{http://mcplots.cern.ch}, and
  a preview of the heavy-ion update at
  \url{http://mcplots-alice.cern.ch}.}.

\begin{figure}[tbp]
  \centering
  \includegraphics[width=.8\linewidth]{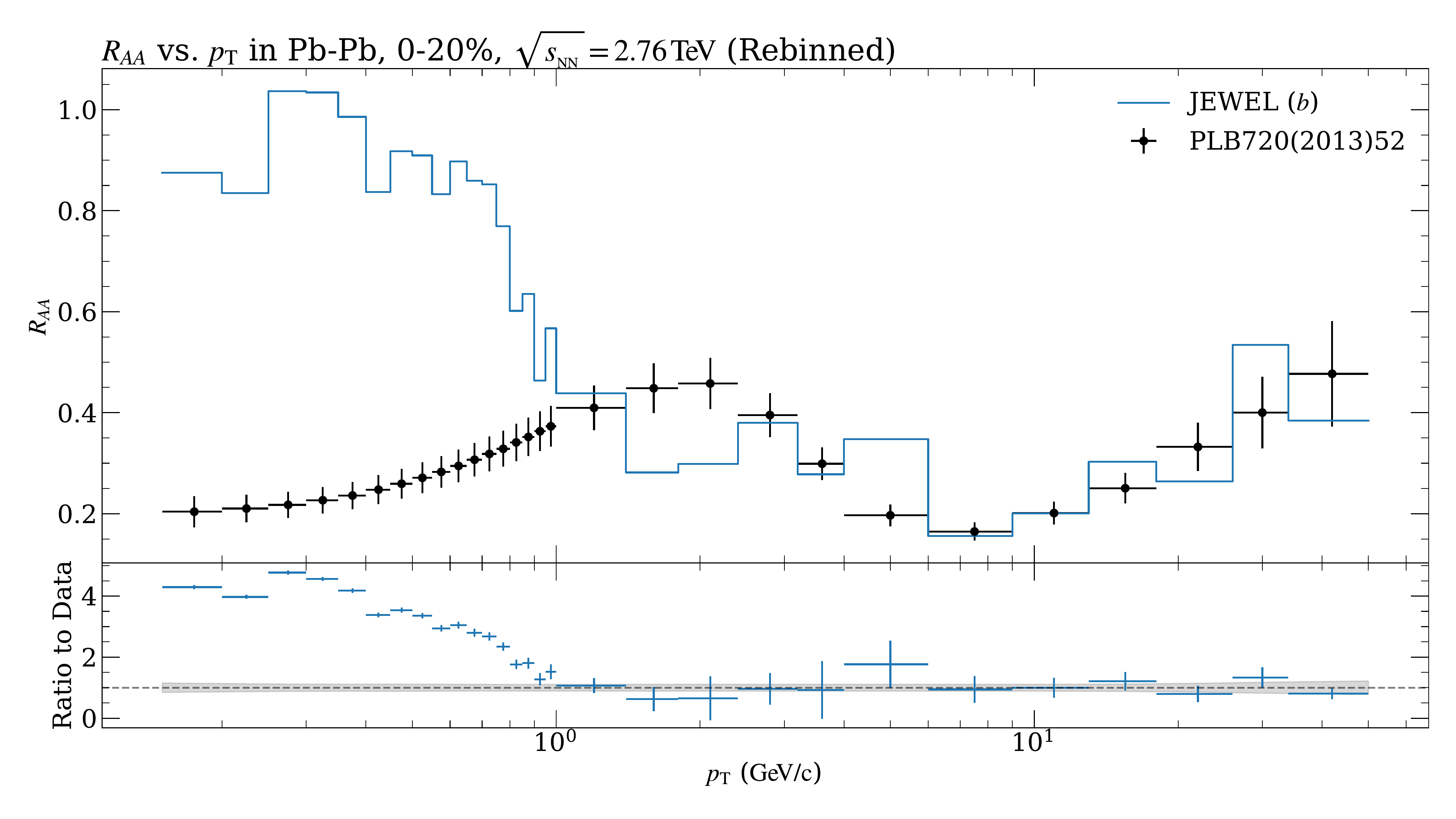}
  \caption{Nuclear-modification factor $\RAA$ measured by
    ALICE~\cite{Abelev:2012hxa} compared to \jewel model
    calculations. The region below $1\,\GeVc$ is not attempted to be
    modelled correctly in \jewel resulting in the large
    discrepancy. Note, that data and MC have been rebinned to allow a
    reasonable comparison.}
  \label{fig:alice-raa-jewel}
\end{figure}

The feature \emph{reentrant finalize} introduced in
Section~\ref{sec:reentrant}, is for example used in the analysis
\texttt{ALICE\_2012\_I1127497}. The main observable is the
nuclear-modification factor $\RAA$ defined as the ratio of the yield
in \AA collisions over \pp collisions scaled by an appropriate factor
$N_\mathrm{coll}$ to accommodate for the trivial scaling of binary
collisions~\cite{Miller:2007ri}:
\begin{equationno}
  \RAA = \frac{
    \left. \mathrm{d}^{2}N_{\mathrm{ch}}/\mathrm{d} \pT \mathrm{d} y
    \right|_{AA}
  }{
    N_{\mathrm{coll}}\left. \mathrm{d}^{2}N_{\mathrm{ch}}/\mathrm{d}
      \pT \mathrm{d} y \right|_{pp}
  }
\end{equationno}
In order to compute this quantity, \rivet is executed three times,
first for a generator with \AA settings, second with \pp settings, and
a third and final time to compute the ratio. A comparison of this
observable to a calculation with the \jewel generator is shown in
Figure~\ref{fig:alice-raa-jewel}, and can be performed with the
following steps, given generated \hepmc~files of \pp and \AA events by
\jewel.

\begin{enumerate}
\item The desired centrality range is set when events are generated in \jewel.
  The centrality value as well as the impact parameter are saved in the HepMC output.
  The usage of the raw centrality values is supported only from HepMC 3.0 onwards (see Section~\ref{sec:centrality}), therefore the impact parameter centrality calibration
  is needed to be applied\footnote{Note, that \jewel produces a single nucleon--nucleon collision instead of a full \AA collision, and therefore the events are marked as being \pp, np, or nn collisions. As the \rivet routine expects \pp or \AA events, the \texttt{ignore-beams} flag needs to be used.} after events have been generated in the full centrality interval (0--100\%):
\begin{verbatim}
rivet aa-sample.hepmc --ignore-beams -a ALICE_2015_PBPBCentrality
    -o calibration.yoda
\end{verbatim}
  
\item The \rivet analysis is run on the \pp sample:

\begin{verbatim}
rivet pp-sample.hepmc --ignore-beams -a ALICE_2012_I1127497 
    -o jewel-pp.yoda
\end{verbatim}
\item Next the \rivet analysis is run on the \AA sample. Note the
  \texttt{beam} option signifying that this is an \AA run. Since the
  \jewel event generator does not supply this information in the
  \hepmc file, but denotes all beams as \pp beams, the analysis
  requires this additional information:

\begin{verbatim}
rivet aa-sample.hepmc --ignore-beams -p calibration.yoda 
    -a ALICE_2012_I1127497:beam=HI:cent=IMP -o jewel-aa.yoda
\end{verbatim}
	\item The two samples must be merged, and the ratio plots constructed, with \texttt{rivet-merge}, which loads histograms from the two runs above, and runs finalize on the preloaded histograms:

\begin{verbatim}
rivet-merge --merge-option beam --merge-option cent 
    jewel-pp.yoda jewel-aa.yoda -o jewel-merged.yoda
\end{verbatim}
		The optional argument \texttt{merge-option} removes the analysis options \texttt{beam} and \texttt{cent} from the heavy ion sample.
	\item The merged sample can be plotted. There is no need to plot the unmerged histograms, as all histograms are contained in the merged file.

\begin{verbatim}
rivet-mkhtml jewel-merged.yoda
\end{verbatim}
\end{enumerate}

\section{Understanding biases in simulation and data}
\label{sec:biases}

Apart from direct validation of a given \mc event generator, the
analysis capabilities of \rivet are useful as tools to reveal inherent
biases in representation of data. In this section the use of the
physics motivated extensions presented in Section~\ref{sec:features}
are exemplified by presenting short analyses of centrality biases
introduced by using different centrality observables for theory and
data (Section~\ref{sec:cent-bias}) and from inclusion of non-flow
effects in flow measurements (Section~\ref{sec:non-flow}). These two
effects are well known, and chosen to show the capabilities of the
framework, rather than presenting an independent physics point.

\subsection{Centrality observables}
\label{sec:cent-bias}

In \pA collisions, such as those performed at the \lhc with \pPb
collisions at $\sNN{5.02}{\TeV{}}$, and recorded by the \atlas
experiment \cite{Aad:2015zza}, the notion of centrality is used as in
\AA collisions, shown in Section~\ref{sec:example}, but without the
clear relation to the physical impact parameter.  Since the physical
impact parameter is well defined and useful from a theoretical point of view (as it can be
used to infer properties of a created QGP phase), the correlation
between modelled impact parameter and measured centrality is worth
having direct access to, as the effects of possible differences
between the two in a measurement is important to understand.

\begin{figure}
  \includegraphics[width=\twofigw]{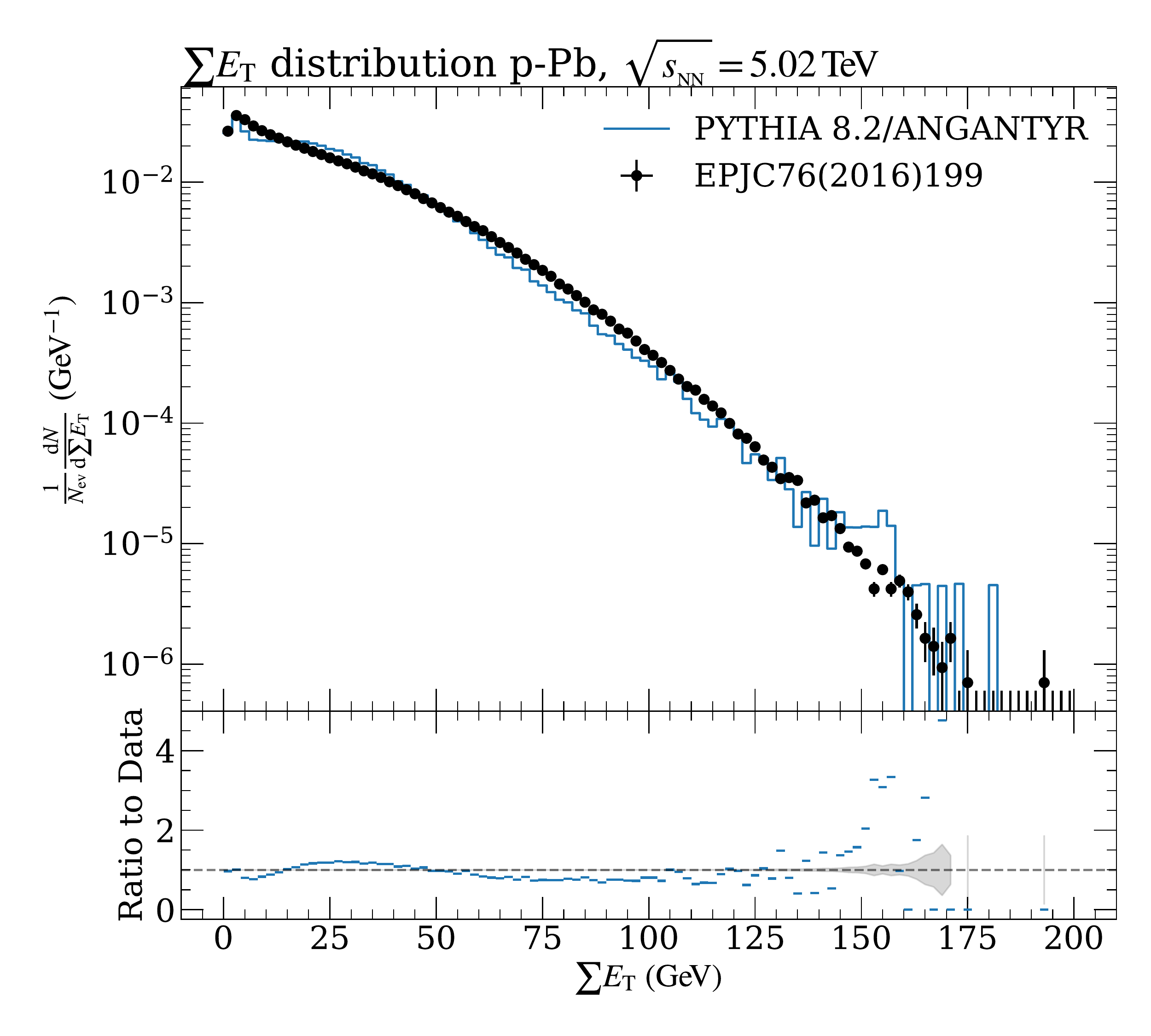}
  \includegraphics[width=\twofigw]{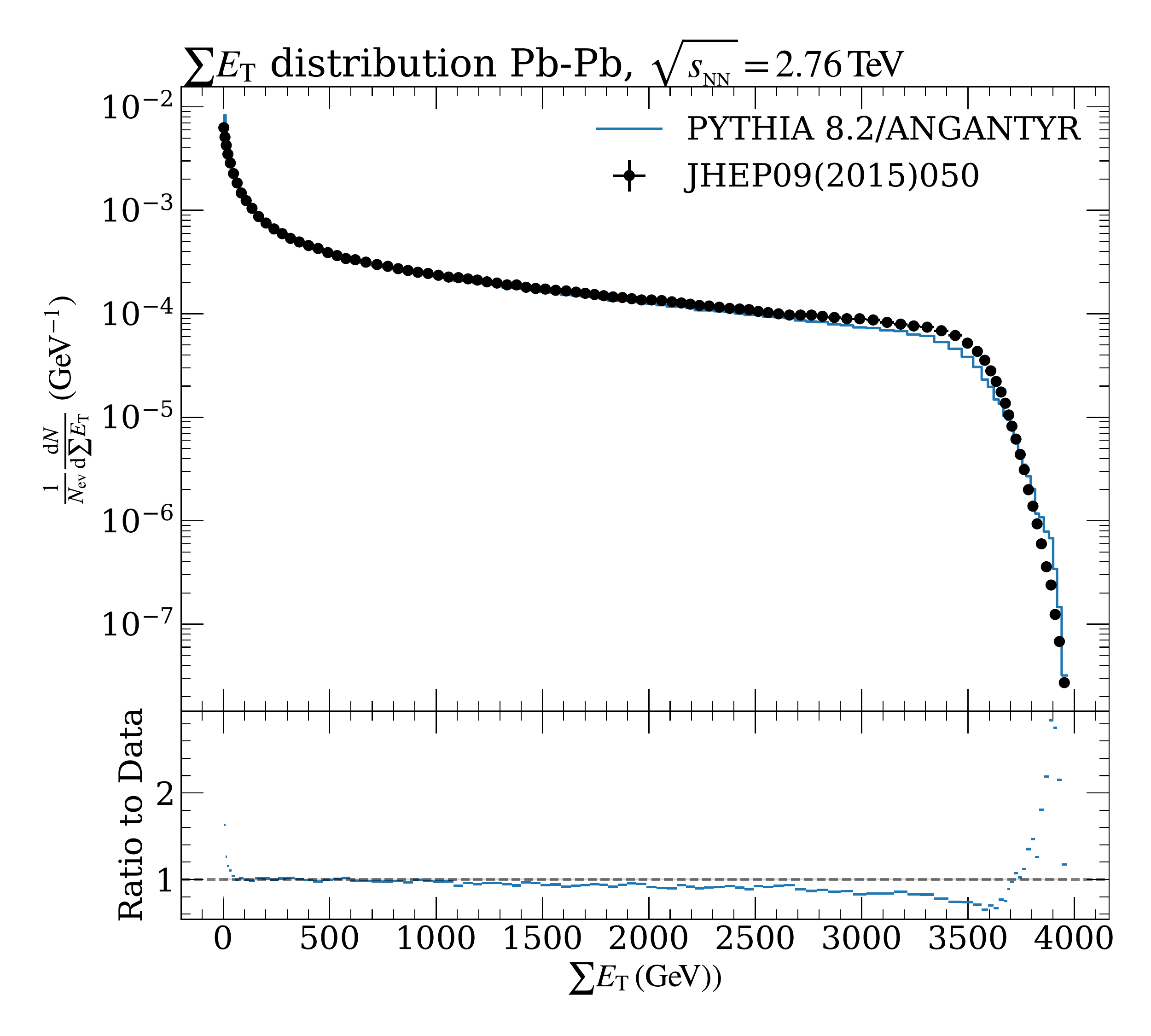}
  \caption{\label{fig:atlas-cent}Centrality measure by \atlas for \pPb
    at $\sNN{5.02}{\TeV{}}$~\cite{Aad:2015zza} (left) and \PbPb at
    $\sNN{2.76}{\TeV{}}$~\cite{Aad:2015wga} (right), compared to
    \pythia. Note that data points are not unfolded, and taken from
    the figures in the paper.}
\end{figure}

The \rivet~centrality implementation offers the possibility to study
such effects directly in specific simulations, by comparing results
from measured centrality to \mc using different centrality
definitions. Consider the centrality measure used by \atlas, shown in
Figure~\ref{fig:atlas-cent}, which is the $\sum \ET$ in the forward
(lead) going direction~\cite{Aad:2015zza}, overlaid with a comparison
to \angantyr. This can be contrasted with the similar
centrality measure used by \atlas in \PbPb collisions at
$\sNN{2.76}{\TeV{}}$~\cite{Aad:2015wga}, which is $\sum \ET$ in both
forward and backward directions. For both centrality measures, it
should be noted that data points are read-off from the
paper. Furthermore the distributions are not unfolded. This adds an
unknown source of uncertainty to the direct comparison to the
centrality, as well as to all derived results, when the experimental
centrality estimate is used.

\begin{figure}
  \includegraphics[width=\twofigw]{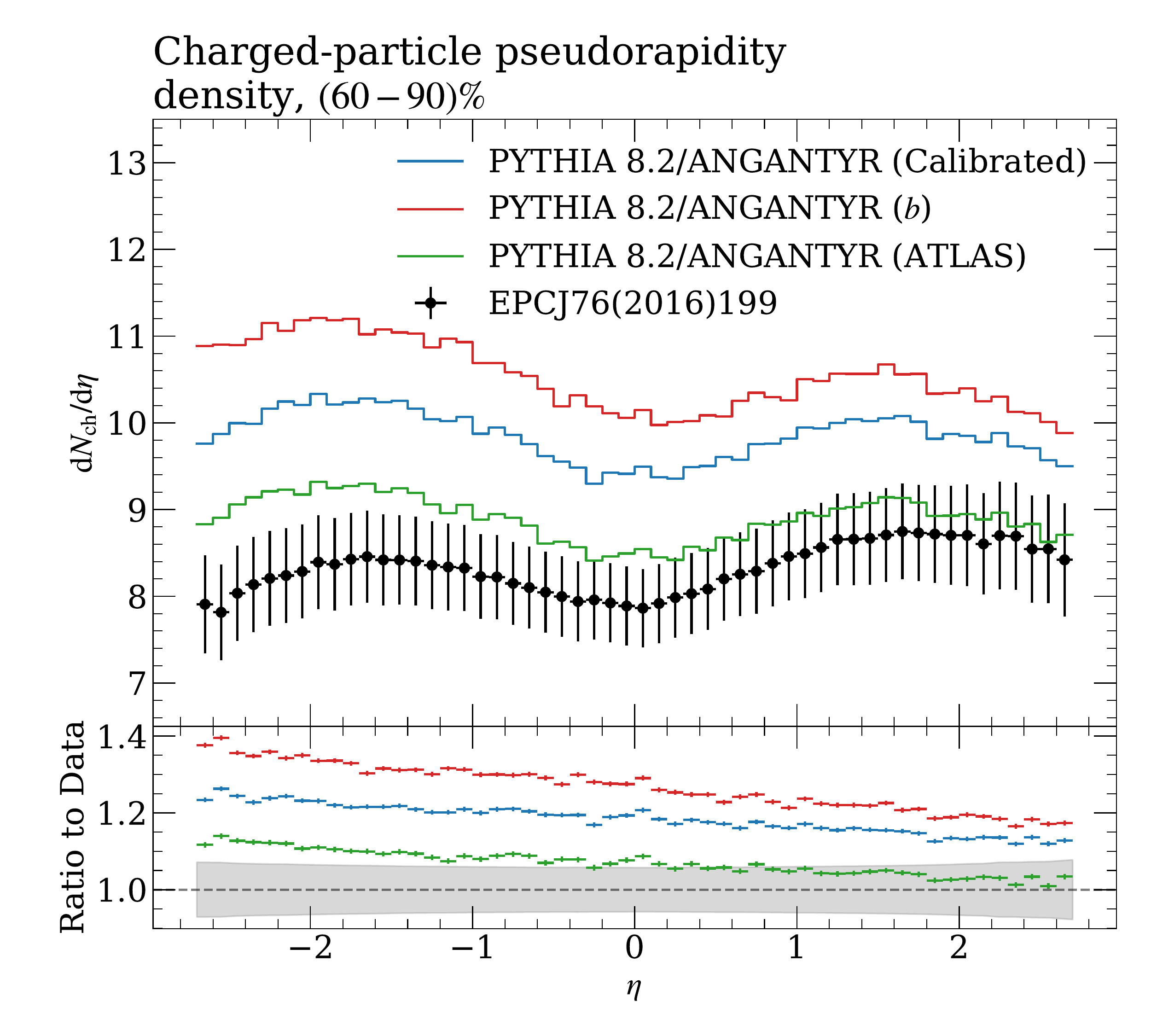}
  \includegraphics[width=\twofigw]{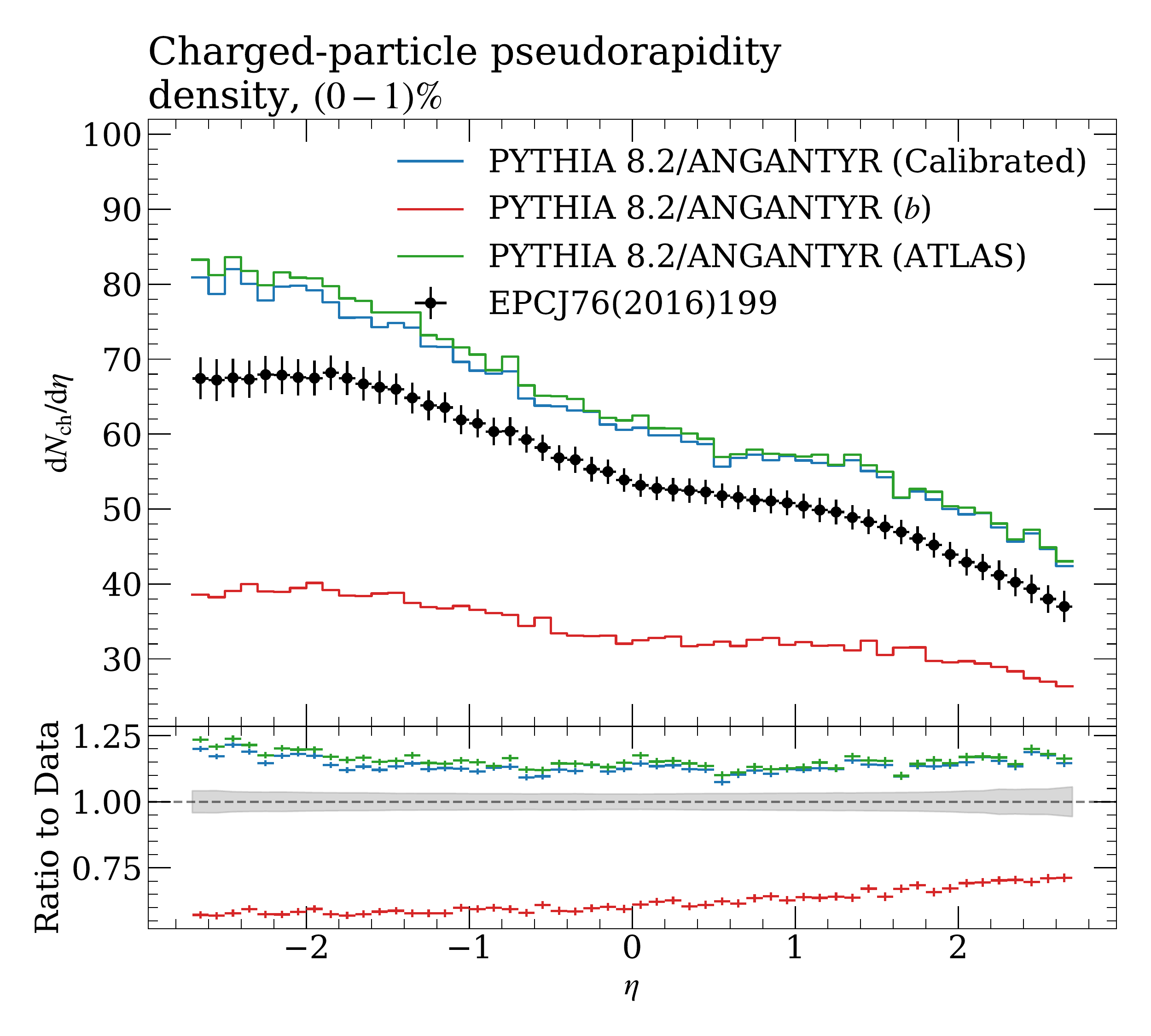}
  \caption{\label{fig:atlas-ppb-mult}Charged-particle pseudorapidity
    density as a function of pseudorapidity in \pPb at
    $\sNN{5.02}{\TeV{}}$ for peripheral (60-90\%) (left) and central
    (0-1\%) (right) events~\cite{Aad:2015wga}, compared to \pythia.}
\end{figure}

In Figure~\ref{fig:atlas-ppb-mult}, the
$\mathrm{d}N_{\mathrm{ch}}/\mathrm{d}\eta$ distributions for the three
different centrality selections, experimental reference calibration
(Experiment, option \texttt{REF}),
generator calibrated (Calibrated, option \texttt{GEN}), and impact
parameter ($b$, option \texttt{IMP}) in \pPb are shown for peripheral
(60-90\%) (left) and central (0-1\%) (right)
events~\cite{Aad:2015zza}. We see a small, but systematic, overestimate of the
charged-particle pseudorapidity density when using
either generator or experimentally calibrated centrality estimators.
The impact parameter based centrality estimator has overall larger deviations
from experimental results (also in
centrality classes not shown here, see ref. \cite{Bierlich:2018xfw} for more comparisons),
but without any clear systematics in the deviations.

\begin{figure}
  \includegraphics[width=\twofigw]{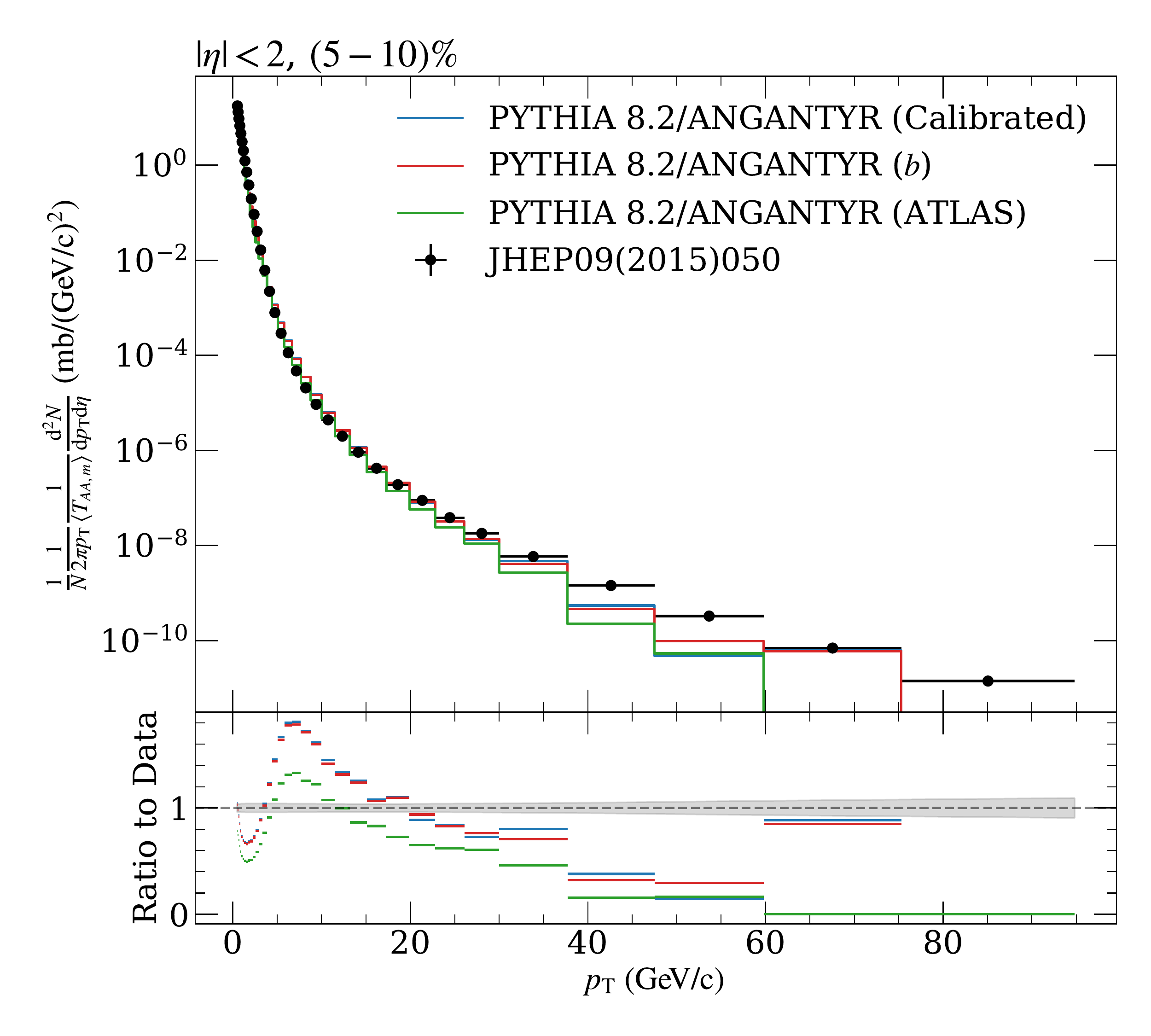}
  \includegraphics[width=\twofigw]{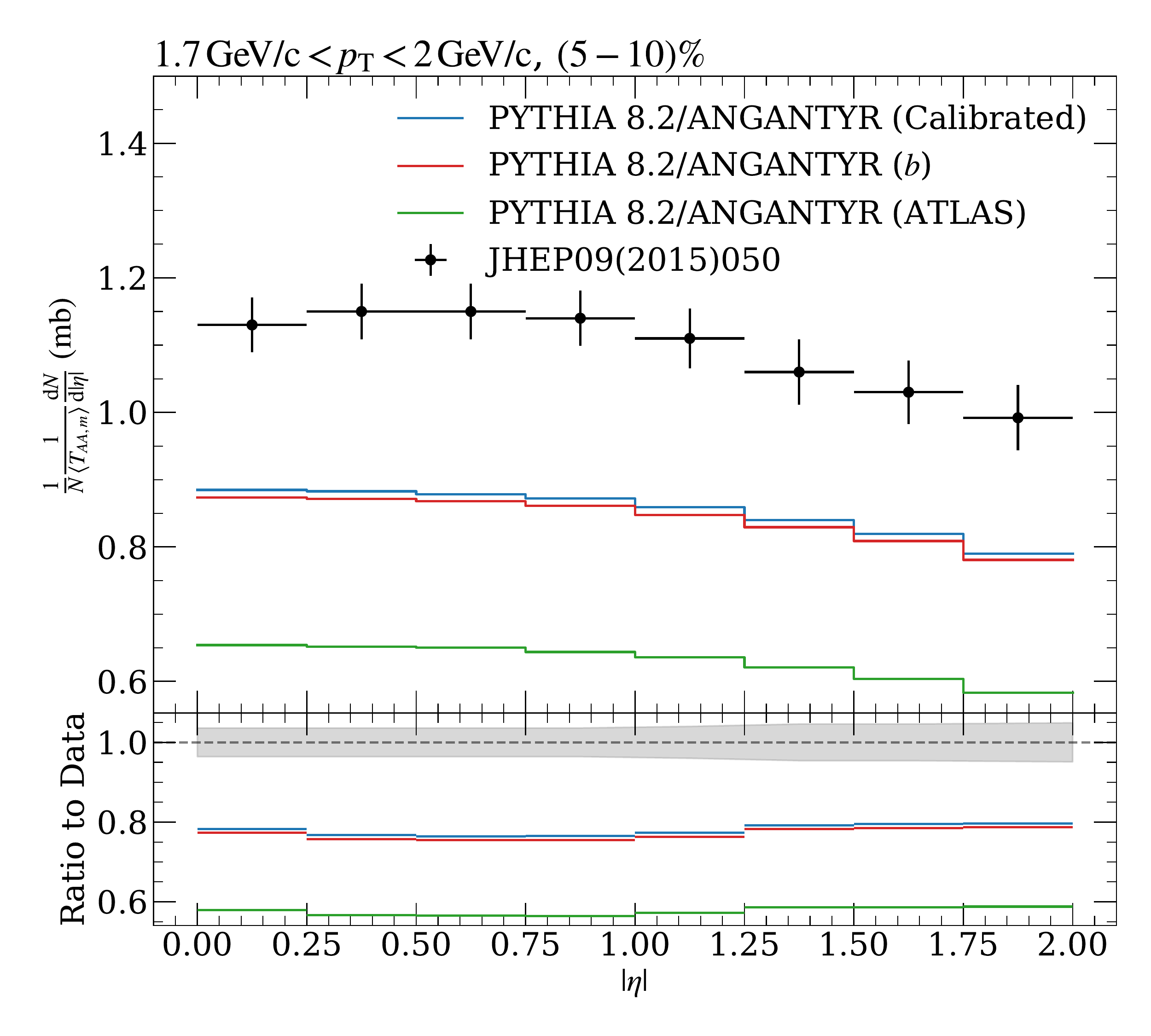}
  \caption{\label{fig:atlas-pbpb-mult}Charged-particle multiplicity as
    function of $\pT$ (left) and $\eta$ for
    $1.7\,\GeVc{} < \pT < 2.0\,\GeVc$ (right) in \PbPb at
    $\sNN{2.76}{\TeV{}}$, in the 5-10\% centrality
    bin~\cite{Aad:2015wga}, compared to \pythia.}
\end{figure}

In Figure~\ref{fig:atlas-pbpb-mult} the charged-particle multiplicity
in \PbPb differential in $\pT$ (left) and $\eta$ for
$1.7\,\GeVc < \pT < 2.0\,\GeVc$ (right) in the 5-10\% centrality bin
are shown~\cite{Aad:2015wga}, again for the three different centrality
estimators. Here, the results from the generated and impact parameter
centrality estimators fully coincides, thus in agreement with the
assumption that the measured centrality (equation~(\ref{eq:n-cent}))
coincides with theoretical centrality (equation~(\ref{eq:b-cent})) in
\AA collisions.  It is also seen that comparison using the measured
experimental centrality measure, deviates from the two others, as
expected.

\subsection{Non-flow effects}
\label{sec:non-flow}

\begin{figure}
  \includegraphics[width=\twofigw]{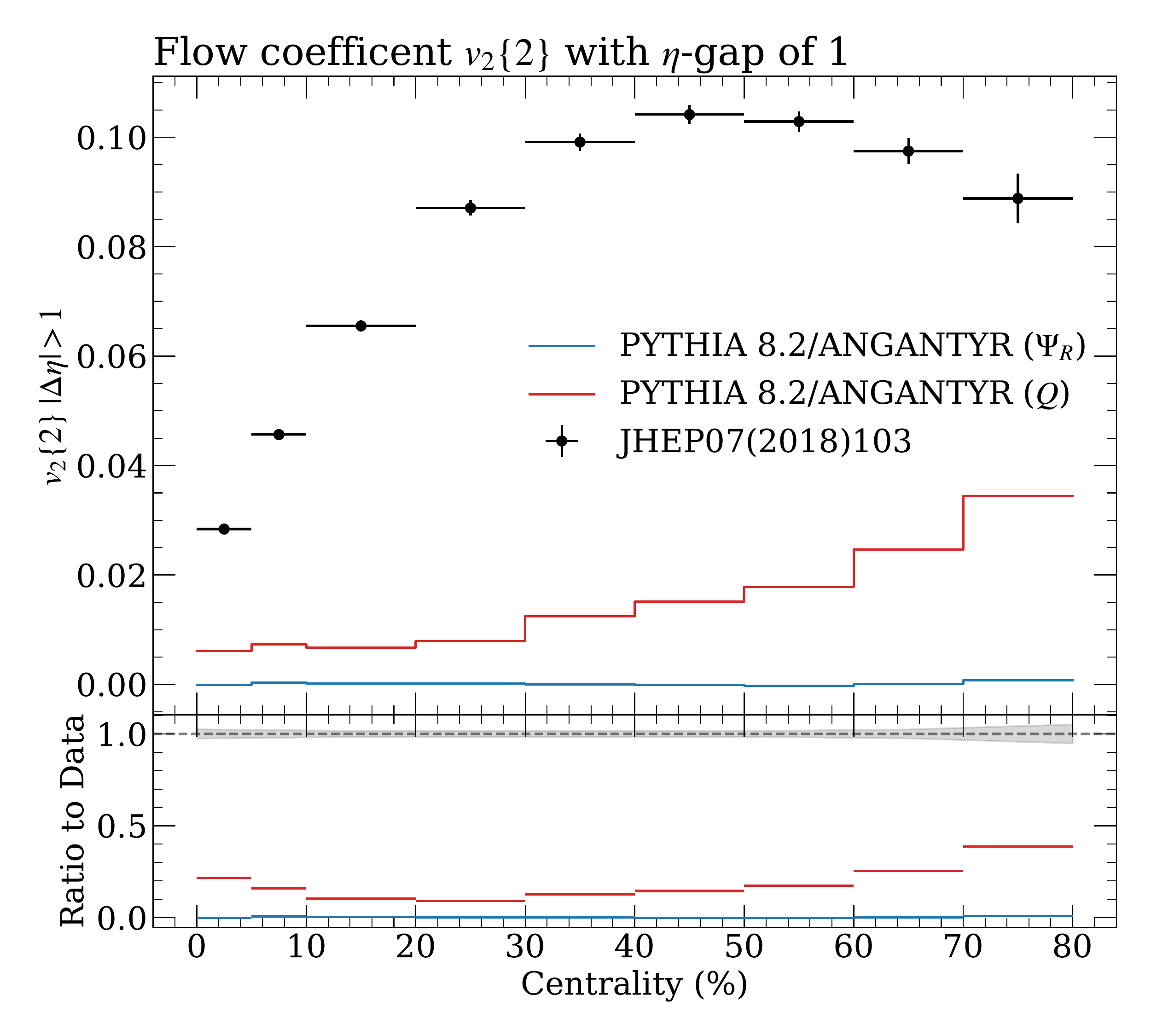}
  \includegraphics[width=\twofigw]{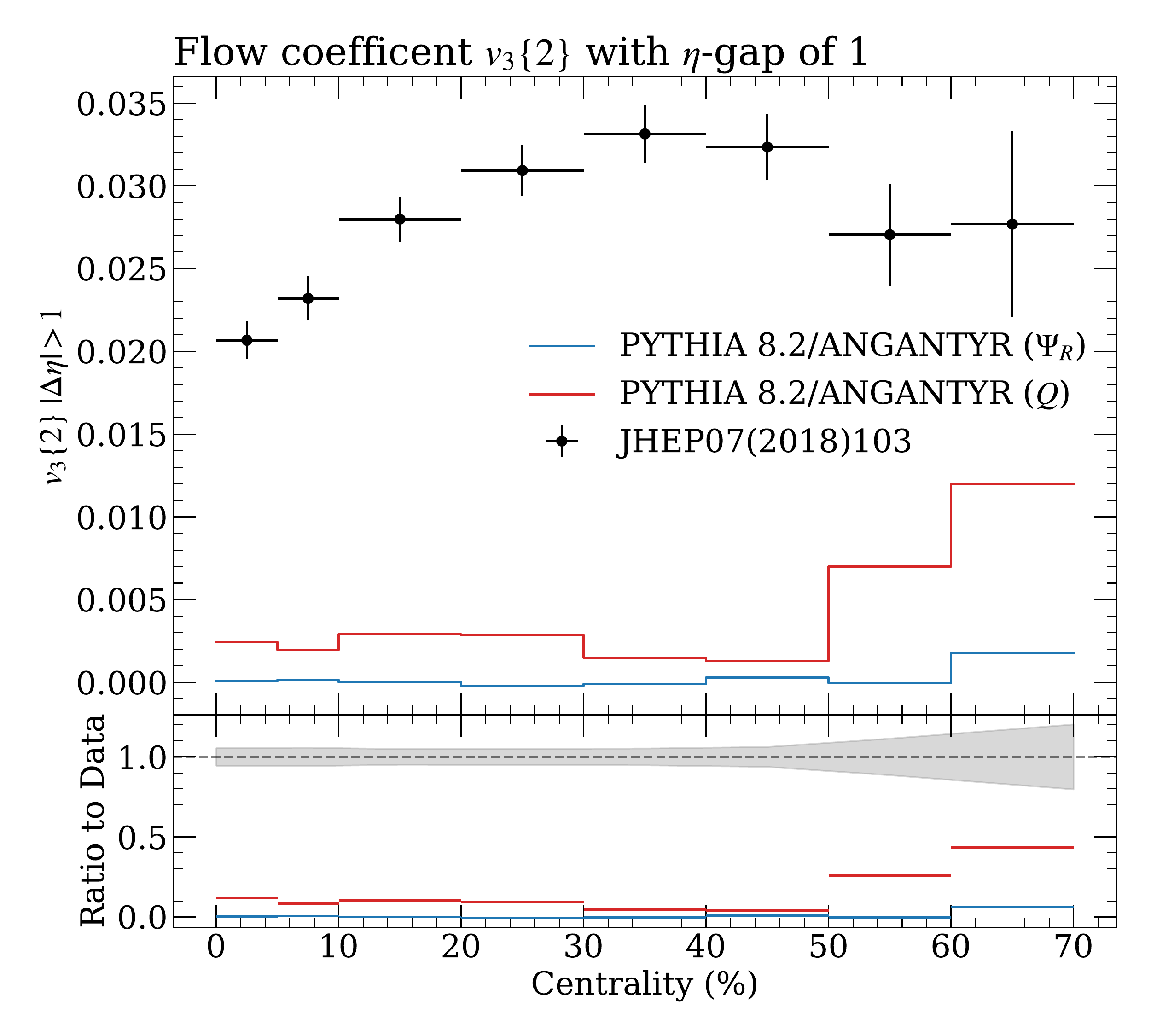}
  \caption{\label{fig:alice-nonflow} Flow coefficients $v_{2}\{2\}$
    (left) and $v_{3}\{2\}$ (right) in \PbPb collisions, at
    $\sNN{2.76}{\TeV{}}$ as a function of centrality, evaluated with
    the event plane method ($\Psi_R$) and the Generic
    Framework ($Q$) implementation in \rivet. Data from
    \alice~\cite{Acharya:2018lmh} compared to \pythia.}
\end{figure}

The $v_{2}$ and $v_{3}$ measured by \alice in \PbPb collisions, at
$\sNN{2.76}{\TeV{}}$ as a function of centrality, is shown in
Figure~\ref{fig:alice-nonflow} together with
comparisons to \angantyr simulations~\cite{Acharya:2018lmh}. The simulations were
obtained using two different approaches. First the event plane method,
where equation~(\ref{eq:flowdefinition}) is evaluated directly, giving
$\langle v_{n} \rangle = \langle
\cos\left(n\left(\phi-\Psi\right)\right)\rangle$. The used value for
the event plane angle $\Psi$\footnote{$\Psi$ is the the angle between
  the impact parameter vector and the beam axis} can be evaluated in
simulations, but not in data. Secondly the Generic Framework (GF)
implementation in \rivet, described in
Section~\ref{sec:generic-framework}, is used to estimate $v_{2}$ and
$v_{3}$ with two-particle correlations as it is done in the
experiment. In the former case, the \pythia simulations are consistent
with 0. This is expected, as the typical mechanisms associated with
the non-flow (jets, resonance decays) are not correlated with the
event plane, and thus any coincidental correlations are averaged
out. On the other hand, this is not true for the simplest case of
two-particle correlations. In particular, back-to-back jets give rise
to strong correlations that are unrelated to the hydrodynamical
evolution of the system. As a result, \pythia~predicts non-zero
$v_{2}$ and $v_{3}$ values calculated with two-particle correlations,
as seen in Figure~\ref{fig:alice-nonflow}. We note that the non-flow
effects are more pronounced for $v_{2}$ and increase for peripheral
\PbPb collisions, where the $\mathrm{d}N_{\mathrm{ch}}/\mathrm{d}\eta$
is lower and therefore the different non-flow sources get less
averaged out. It is noted that the non-flow contribution is
here shown to be sizeable for $v_2\{2\}$ even with a required minimum
separation in $\eta$ between the two correlated particles.

\section{Collective effects in \pp collisions}

\begin{figure}
  \includegraphics[width=\twofigw]{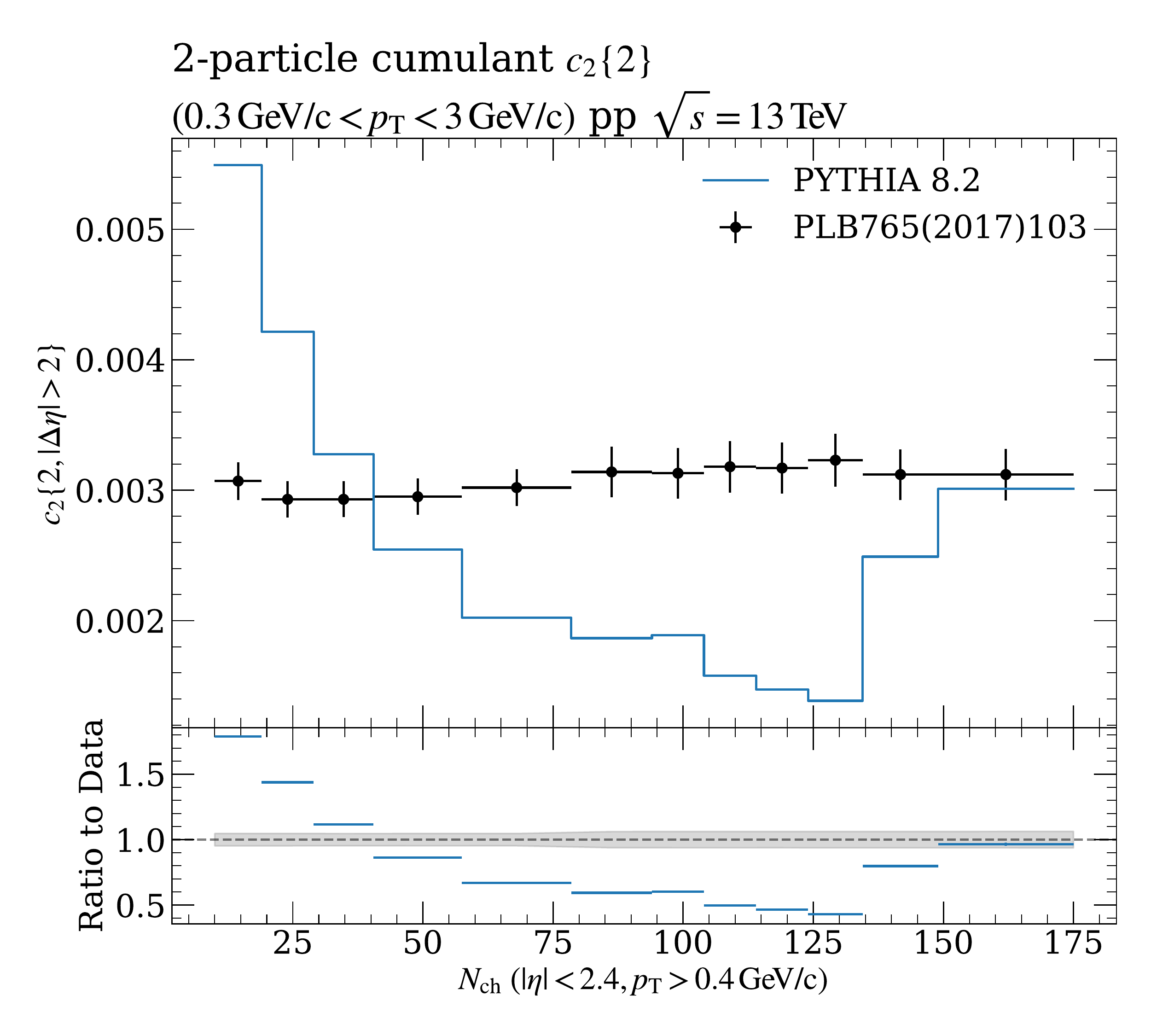}
  \includegraphics[width=\twofigw]{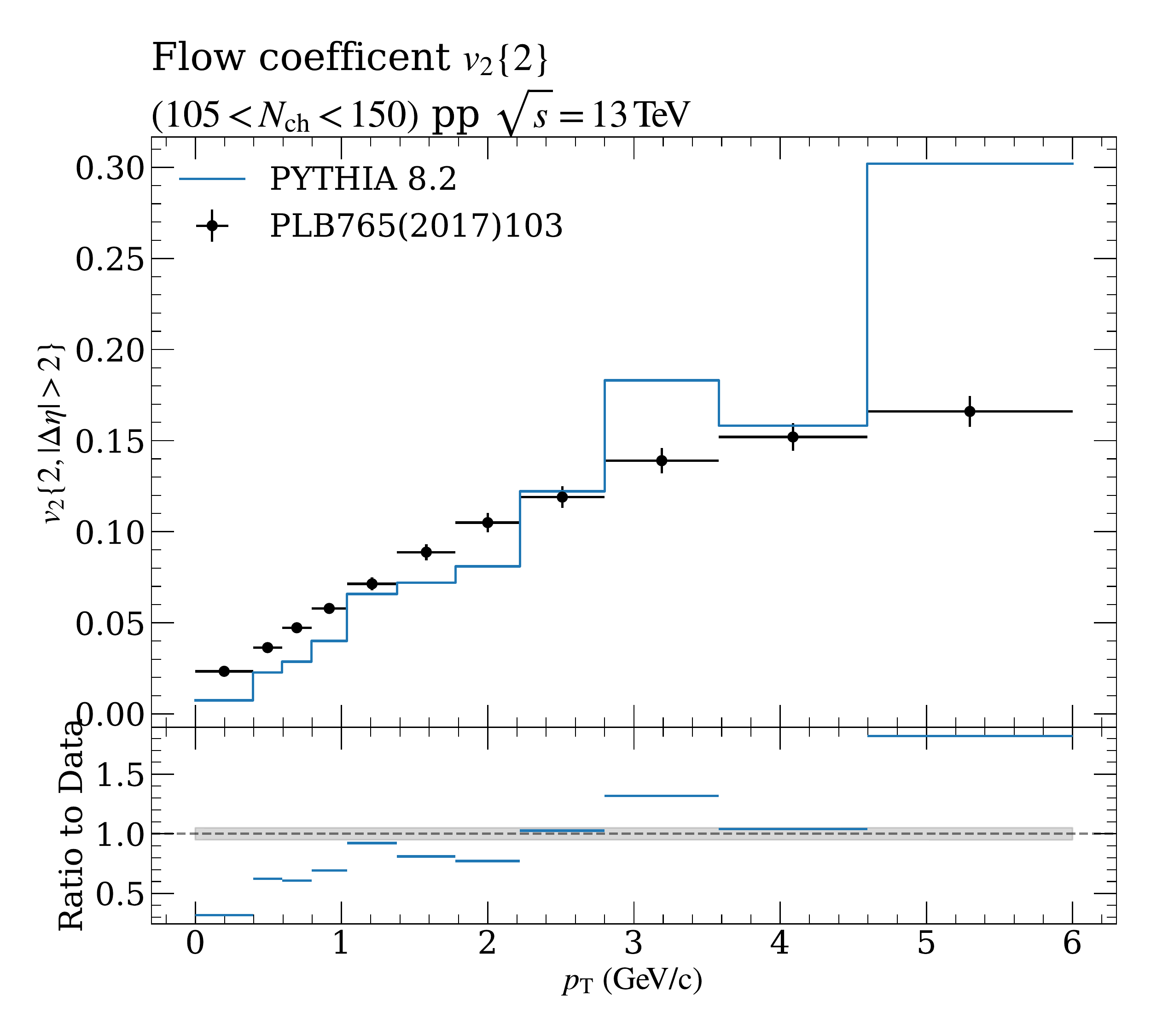}
  \caption{\label{fig:cms-mult}Example flow observables from \pp
    collisions~\cite{Khachatryan:2016txc}, compared to
    simulation with \pythia. Analysis performed with the
    implementation of the Generic Framework for flow measurements
    \cite{Bilandzic:2013kga} presented in Section~\ref{sec:generic-framework}.}
\end{figure}

The discovery of heavy-ion-like effects in \pp collisions has spurred
a lot of interest from the proton-proton \mc authors to introduce
models for such effects
\cite{Bierlich:2017vhg,Gieseke:2017clv,Duncan:2018gfk,Gieseke:2018gff,Bierlich:2015rha,
Bierlich:2014xba,Christiansen:2015yqa,Fischer:2016zzs}.
The inclusion of heavy-ion features in \rivet~will aid this
development, as such features can be used directly in new analyses for
\pp as well. As an example, the seminal results from the \cms
experiment on two- and multi-particle correlations
\cite{Khachatryan:2016txc} and from the \alice experiment on
strangeness enhancement~\cite{ALICE:2017jyt} has been implemented in
\rivet~analyses \texttt{CMS\_2017\_I1471287} and
\texttt{ALICE\_2016\_I1471838}.  Results from the former shown in
Figure~\ref{fig:cms-mult}.

The analysis is implemented with the flow framework presented in
Section~\ref{sec:generic-framework}. In Figure~\ref{fig:cms-mult} the
cumulant $c_2\{2\}$ as a function of charged-particle multiplicity
(left, equation~(\ref{eq:c22})), and the corresponding flow
coefficient as a function of transverse momentum (right) are shown.

\section{Conclusion and future work}

Extensions to the \rivet~framework to facilitate comparisons between
\mc{event generators} and data have been presented. The
\rivet~framework is a standard tool for preservation of \lhc analyses
of proton--proton collisions, and with the presented extensions, the
framework is ready to start taking a similar role for heavy-ion
physics.

There are, however, areas where the updated framework is foreseen to
be further extended in order to allow for better comparisons to jet
physics studies, reproduction of global fits of QGP properties, as
well as allowing for comparison of simulation tools developed for
heavy-ion physics.

One avenue of further development is jet observables. \rivet includes
an interface to the popular FastJet package~\cite{Cacciari:2011ma} for
jet reconstruction, which already includes some functionality for
sub-jet observables geared towards heavy-ion
collisions~\cite{Andrews:2018jcm}. A more substantial problem for
preservation of jet analyses is that of handling the large
``underlying event'' of a heavy-ion collision. A preliminary \rivet
implementation to perform background subtraction and QGP
back-reactions for specific models is
available~\cite{KunnawalkamElayavalli:2017hxo}, but no truly model
independent prescription exist. Due to large differences between
models, it may even be such that experimental background subtraction
cannot be fully reproduced for models unless the model also attempts
to describe the full underlying event. In such case a better option
might be to publish data without background subtraction for model
comparisons.

Another future direction involves expanding the possibility for
post-processing beyond what re-entrant finalization
allows. Specifically to add the possibility to perform fits used in
experimental analyses, such as for example reinterpretation in terms
of simple thermal models like Boltzmann--Gibbs Blast Wave for
\AA~\cite{Schnedermann:1993ws} or Lévy--Tsallis for
\pp~\cite{Tsallis:1987eu,Abelev:2006cs}. A technically similar
development for high energy physics is realized in the \textsc{Contur}
package~\cite{Butterworth:2016sqg}, which constrains BSM physics using
Standard Model measurements implemented in \rivet. Reinterpretation of
heavy-ion analyses could be performed in a technically similar way,
possibly also as an additional software package. Such an extension
could also cater to theoretical models not suitable for direct
event-by-event analysis.

\section{Acknowledgements}
We thank the Discovery Center at the Niels Bohr Institute for hosting
a successful workshop in August 2018, consolidating the efforts
described in this paper, and in particular the NBI \alice group for providing
organizational support. The workshop was further supported by the
Marie Sk\l odowska-Curie Innovative Training Network MCnetITN3 (grant
agreement 722104).

\noindent The authors would further like to acknowledge support from individual
funding bodies:
\begin{itemize}[leftmargin=2.5ex]
\item[--] CB was supported by the Swedish Research Council, contract number
2017-003.
\item[--] AB was supported by The Royal Society via University Research Fellowship grant
UF160548.
\item[--] COR and LL were supported in part by the European Research Council
(ERC) under the European Union's Horizon 2020 research and innovation
program (grant agreement No 668679), and in part
by the Swedish Research Council, contract number 2016-05996.
\item[--] LL was supported in part by the Swedish Research Council, contract number
2016-03291, and the Knut and Alice Wallenberg foundation, contract no
2017.0036.
\item[--] PaKi acknowledges support by the BMBF under grant number
05H18VKCC1.
\end{itemize}

\bibliographystyle{iopart-num}
\bibliography{paper,otherbib}
\appendix
\section{Analyses with heavy ion features}
This appendix contains a list of all \rivet analyses implementing one
or more heavy ion features, or heavy ion beams, at the time of
writing, shown in table \ref{table:analyses}. It includes a mention of
features used, as a handy guide to anyone looking for inspiration to
implement a specific feature into their analyses. An up-to-date list
of all \rivet~analyses can be found at
\texttt{https://rivet.hepforge.org/}.  \vspace{0.5cm}
\begin{table}
  \caption{\label{table:analyses} All \rivet analyses implementing one
    or more heavy ion features, or heavy ion beams. An up-to-date list
    can be found at \texttt{https://rivet.hepforge.org}.}
\begin{adjustbox}{width=\textwidth,center}
\begin{tabular}{lcccr}
  \toprule
  Analysis name & System & $\nsNN{}$ & Heavy ion features & Reference \\
  \midrule
  \verb:ALICE_2010_I880049:  & \PbPb & $2.76\,\TeV$ & centrality, primary particles, heavy ion container & \cite{Aamodt:2010cz} \\
  \verb:ALICE_2012_I930312:  & \PbPb & $2.76\,\TeV$ & centrality, heavy ion container, re-entrant finalize & \cite{Aamodt:2011vg} \\
  \verb:ALICE_2012_I1127497: & \PbPb & $2.76\,\TeV$ & centrality, heavy ion container, re-entrant finalize & \cite{Abelev:2012hxa} \\
  \verb:ALICE_2012_I1126966: & \PbPb & $2.76\,\TeV$ & centrality, primary particles & \cite{Abelev:2012wca} \\
  \verb:ALICE_2013_I1225979: & \PbPb & $2.76\,\TeV$ & centrality, primary particles & \cite{Abbas:2013bpa} \\
  \verb:ALICE_2014_I1243865: & \PbPb & $2.76\,\TeV$ & centrality, primary particles & \cite{ABELEV:2013zaa} \\
  \verb:ALICE_2014_I1244523: & \pPb  & $5.02\,\TeV$ & centrality, primary particles & \cite{Abelev:2013haa} \\
  \verb:ALICE_2016_I1394676: & \PbPb & $2.76\,\TeV$ & centrality, primary particles & \cite{Adam:2015kda} \\
  \verb:ALICE_2016_I1419244: & \PbPb & $5.02\,\TeV$ & centrality, generic framework & \cite{Adam:2016izf} \\
  \verb:ALICE_2016_I1507090: & \PbPb & $5.02\,\TeV$ & centrality, primary particles & \cite{Adam:2016ddh} \\
  \verb:ALICE_2016_I1507157: & \pp   & $7\,\TeV$    & event mixing                  & \cite{Adam:2016iwf} \\
  \verb:ATLAS_2015_I1360290: & \PbPb & $2.76\,\TeV$ & centrality                    & \cite{Aad:2015wga} \\
  \verb:ATLAS_2015_I1386475: & \pPb  & $5.02\,\TeV$ & centrality                    & \cite{Aad:2015zza} \\
  \verb:ATLAS_PBPB_CENTRALITY: & \PbPb & $2.76\,\TeV$ & centrality                  & \cite{Aad:2015wga} \\
  \verb:ATLAS_pPb_Calib:     & \pPb  & $5.02\,\TeV$ & centrality                    & \cite{Aad:2015zza} \\
  \verb:BRAHMS_2004_I647076: & \AuAu & $200\,\GeV$  & centrality, primary particles & \cite{Bearden:2004yx} \\
  \verb:CMS_2017_I1471287:   & \pp   & $13\,\TeV$   & generic framework             & \cite{Khachatryan:2016txc} \\
  \verb:LHCF_2016_I138587:   & \pPb  & $5.02\,\TeV$ & -                             & \cite{Adriani:2015iwv} \\
  \verb:STAR_2016_I1414638:  & \AuAu & $7.7-200\,\GeV$ & centrality                 & \cite{Adamczyk:2016exq} \\
  \verb:STAR_2017_I1510593:  & \AuAu & $7.7-39\,\GeV$ & centrality             & \cite{Adamczyk:2017iwn} \\
  \bottomrule
\end{tabular}
\end{adjustbox}
\end{table}

\end{document}